\documentclass[journal,twoside,web]{mainclass}
\usepackage{geometry}
\usepackage{setspace,lineno}

\usepackage{amsmath,amssymb,amsfonts}
\usepackage{amsthm}
\usepackage{mathrsfs}

\usepackage{graphicx}
\usepackage{float}
\usepackage{multirow}
\usepackage{booktabs}
\usepackage[table]{xcolor}
\usepackage{colortbl}
\usepackage{multicol}
\usepackage{tabularx}
\usepackage{tablefootnote}
\usepackage[table,xcdraw]{xcolor}

\usepackage{algorithm}
\usepackage{algpseudocode}

\usepackage{listings}

\usepackage{cite}
\usepackage{manyfoot}

\usepackage{xcolor}
\usepackage{textcomp}
\usepackage{pifont}
\usepackage{soul}
\usepackage{tikz}

\usepackage[table]{xcolor}
\definecolor{customgreen}{HTML}{00B050}
\definecolor{orcidgreen}{HTML}{A6CE39}
\newcommand{\cmark}{\textcolor{customgreen}{\ding{51}}} 
\newcommand{\xmark}{\textcolor{red}{\ding{55}}}

\newcommand{\orcidicon}[1]{%
  \textsuperscript{%
    \href{https://orcid.org/#1}{%
      \begin{tikzpicture}[baseline=-0.1em]
        \fill[orcidgreen] (0,0) circle (1.5ex);
        \node[white, scale=0.8, font=\bfseries\sffamily] at (0,0) {iD};
      \end{tikzpicture}%
    }%
  }%
}

\makeatletter
\let\NAT@parse\undefined
\makeatother

\usepackage[colorlinks=true, linkcolor=blue, citecolor=blue, urlcolor=blue]{hyperref}

\begin{document}

\author{Arefin Ittesafun Abian\orcidicon{0009-0003-4451-0838}, Ripon Kumar Debnath\orcidicon{0009-0009-3567-8288}, Md. Abdur Rahman\orcidicon{0009-0004-3097-8576}, Mohaimenul Azam Khan Raiaan\orcidicon{0009-0006-4793-5382}, Md Rafiqul Islam\orcidicon{0000-0001-7209-3881}, Asif Karim\orcidicon{0000-0001-8532-6816}, Reem E. Mohamed\orcidicon{0000-0002-6992-6608}, Sami Azam\orcidicon{0000-0001-7572-9750}

\thanks{Manuscript Submitted 16\textsuperscript{th} July, 2025}

\thanks{(Corresponding author: Sami Azam)}

\thanks{This work did not involve human subjects or animals in its research.}

\thanks{Arefin Ittesafun Abian and Ripon Kumar Debnath contributed equally and are joint-first authors. They are affiliated with the Department of Computer Science and Engineering, United International University, Dhaka, Bangladesh, along with Md. Abdur Rahman Bangladesh (e-mail: aabian191042@bscse.uiu.ac.bd, rdebnath192071@bscse.uiu.ac.bd, mrahman202260@bscse.uiu.ac.bd).}
\thanks{Mohaimenul Azam Khan Raiaan is with the Faculty of Science and Technology, Charles Darwin University, Casuarina, NT 0909, Australia, and also with the Department of Computer Science and Engineering, United International University, Dhaka, Bangladesh (e-mail: mohaimenulazamkhan.raiaan@cdu.edu.au).}
\thanks{Md Rafiqul Islam, Asif Karim and Sami Azam are with the Faculty of Science and Technology, Charles Darwin University, Casuarina, 0909, NT, Australia (e-mail: mdrafiqul.islam@cdu.edu.au, asif.karim@cdu.edu.au,  Sami.Azam@cdu.edu.au).} 
\thanks{Reem E. Mohamed is with the Faculty of Science and Information Technology, Charles Darwin University, Sydney, NSW, Australia (e-mail: reem.sherif@cdu.edu.au).}}

\title{HANS-Net: Hyperbolic Convolution and Adaptive Temporal Attention for Accurate and Generalizable Liver and Tumor Segmentation in CT Imaging}

\markboth{IEEE Transactions on Radiation and Plasma Medical Sciences, Vol. 00, No. 0, July 2025}
{Abian, Debnath \MakeLowercase{\textit{et al.}}: IEEE Transactions on Radiation and Plasma Medical Sciences}

\maketitle

\begin{abstract}

 Accurate liver and tumor segmentation on abdominal CT images is critical for reliable diagnosis and treatment planning, but remains challenging due to complex anatomical structures, variability in tumor appearance, and limited annotated data. To address these issues, we introduce \textbf{H}yperbolic-convolutions \textbf{A}daptive-temporal-attention with \textbf{N}eural-representation and \textbf{S}ynaptic-plasticity \textbf{Net}work (HANS-Net), a novel segmentation framework that synergistically combines hyperbolic convolutions for hierarchical geometric representation, a wavelet-inspired decomposition module for multi-scale texture learning, a biologically motivated synaptic plasticity mechanism for adaptive feature enhancement, and an implicit neural representation branch to model fine-grained and continuous anatomical boundaries. Additionally, we incorporate uncertainty-aware Monte Carlo dropout to quantify prediction confidence and lightweight temporal attention to improve inter-slice consistency without sacrificing efficiency. Extensive evaluations of the LiTS dataset demonstrate that HANS-Net achieves a mean Dice score of 93.26\%, an IoU of 88.09\%, an average symmetric surface distance (ASSD) of 0.72 mm, and a volume overlap error (VOE) of 11.91\%. Furthermore, cross-dataset validation on the AMOS 2022 dataset obtains an average Dice of 85.09\%, IoU of 76.66\%, ASSD of 19.49 mm, and VOE of 23.34\%, indicating strong generalization across different datasets. These results confirm the effectiveness and robustness of HANS-Net in providing anatomically consistent, accurate, and confident liver and tumor segmentation.
 
\end{abstract}


\section{Introduction}
\IEEEPARstart{T}{he} liver is a vital glandular organ that performs functions of digestion, metabolism, detoxification, and immunity \cite{zhan2023three}. Liver cancer ranks among the five most frequently diagnosed cancers and constitutes the fourth leading cause of cancer-related mortality worldwide, with hepatocellular carcinoma (HCC) being the most prevalent form \cite{zhang2021deeprecs}. Precise segmentation of liver tumors is essential in the diagnosis and radiation therapy of HCC. According to the World Health Organization (WHO), cancer accounted for 8.8 million deaths in 2015, with 788,000 specifically attributed to liver cancer \cite{rahman2025automatic,minaee2021image}. With increased public health awareness, the need for effective liver cancer diagnosis and treatment has intensified \cite{wang2024sbcnet, azad2024medical}. Historically, specialists have manually identified liver tumors, an accurate, yet tedious and labor-intensive process prone to errors under substantial workload \cite{wang2024sbcnet, conze2023current}. The development of computer-aided diagnostic systems has introduced intelligent, automated, or partially automated approaches to the analysis of medical images, significantly improving precision, effectiveness, and consistency in the detection and segmentation of tumors. \cite{yang2025multi, liu2021localised, conze2023current}.

Various image processing and machine learning techniques have been developed to improve liver tumor segmentation and support early cancer intervention \cite{gul2022deep, minaee2021image, azad2024medical}. Among these methodologies, attention-based and multi-branch models have earned significant interest, integrating sophisticated architectures such as S2DA-Net \cite{liu2024s2da}, EG-UNETR \cite{cheng2024eg}, Generalized U-Net \cite{dj2024liver}, and ConvNeXt-2U \cite{chen2024convnext}. These models are structured to encapsulate essential semantic features in various imaging phases, increasing both the precision and the robustness of feature extraction \cite{dj2024liver}. Currently, other innovative strategies have taken advantage of multiphase fusion and multiscale representation techniques, such as  PGC-Net \cite{you2024contour} and camouflaged feature mining modules \cite{yang2025multi}, which collectively advance the automated identification of complex anatomical structures, encompassing both liver and tumor regions \cite{wang2024sbcnet}. 

Recent developments in liver tumor segmentation have made progress, yet several challenges persist in comprehensively addressing the complexity of the task. Current liver and tumor segmentation studies frequently encounter limitations in the effective management of complex and heterogeneous tumor-liver anatomies, resulting in only moderate accuracy levels. These models depend on flattened representations, which hinders their ability to accurately capture the complex spatial organization and progression of tumors across 3D slices \cite{you2024contour}. Furthermore, they are often incapable of segmenting tumors and livers simultaneously \cite{hu2025msml}. Most methodologies overlook the enhancement of inter-slice consistency without compromising efficiency. They also do not improve adaptive features and construct continuous and fine-grained anatomical boundaries \cite{yang2025multi, dj2024liver}. Moreover, they typically lack the strategies of temporal attention necessary to capture sequential relationships in volumetric scans and face challenges in resolving fine-grained tumor boundaries due to insufficient detail encoding \cite{wang2024sbcnet}. Finally, the absence of uncertainty estimation prevents the scope of quantifying the confidence in prediction \cite{kuang2024adaptive}. 

The reliance on flattened representations in existing liver tumor segmentation models limits their ability to model complex anatomies, maintain inter-slice consistency, and construct fine-grained anatomical boundaries. This motivated us to develop HANS-Net (Hyperbolic-convolutions Adaptive-temporal-attention with Neural-representation and Synaptic-plasticity Network), a novel segmentation framework designed to jointly segment both tumors and liver tissues with improved anatomical coherence and confidence estimation. While individual components of our architecture have seen use in various domains, to the best of our knowledge, no prior work has jointly applied them to medical image segmentation. For instance, hyperbolic learning methods have been explored for representation learning and graph neural networks \cite{liu2024deephgcn}, but their application to dense 2D or 3D segmentation remains largely unexplored. Similarly, biologically inspired mechanisms such as synaptic plasticity are rarely integrated into medical imaging models; existing segmentation architectures (e.g., U-Net, TransUNet, EG-UNETR) do not employ plasticity-based adaptive modulation. Furthermore, while recent efforts have investigated implicit neural representations \cite{zhang2022implicit}, temporal attention \cite{shi2020spatial}, and uncertainty modeling independently, none have unified them in a single end-to-end system.

HANS-Net addresses these shortcomings through a comprehensive design. It employs hyperbolic convolutions to represent anatomical hierarchies in a non-Euclidean space, enabling accurate modeling of tumor complexity, and a wavelet-inspired decomposition module captures multiscale information. While 3D volumetric approaches naturally preserve spatial continuity, they are computationally intensive and often require extensive annotated data. In contrast, 2D slice-based methods offer computational efficiency and can effectively capture inter-slice relationships when equipped with appropriate mechanisms. Accordingly, HANS-Net integrates adaptive temporal attention (ATA) along with multi-view fusion, making the approach particularly suitable for resource-constrained clinical settings while maintaining competitive segmentation performance. A synaptic plasticity module enhances feature adaptivity, and an implicit neural representation branch improves boundary smoothness. Finally, Monte Carlo dropout is used to quantify uncertainty in the predictions. HANS-Net achieves superior performance (Dice: 93.26\%), surpassing state-of-the-art (SOTA) accuracy and demonstrating strong cross-dataset generalization (Dice: 85.09\% on AMOS 2022), highlighting its robustness and potential for real-world clinical deployment.

The main contributions of this work are as follows.
\begin{itemize}
    \item Introduces HANS-Net, the first liver tumor segmentation framework that unites hyperbolic geometry and implicit neural representations for continuous, hierarchy-aware learning in curved space.
    \item Proposes a learnable wavelet-inspired decomposition module that captures both global liver contours and fine tumor textures, enabling efficient multiscale analysis without heavy computation.
    \item Integrates a bio-inspired synaptic plasticity mechanism that adaptively adjusts feature weights based on activation dynamics, promoting tumor-focused learning without extra supervision.
    \item Employs a coordinate-based implicit MLP branch with positional encoding to refine segmentation outputs, generating anatomically plausible tumor boundaries at continuous resolution.
    \item Enhances clinical trust via pixel-wise uncertainty estimation using Monte Carlo dropout and improves inter-slice consistency with lightweight temporal attention mimicking 3D context.
\end{itemize}

The remainder of this paper is organized as follows. Section \ref{related_works} reviews the related work conducted in this area. Section \ref{method} details our proposed HANS-Net model and its component designs. Section \ref{experiment} describes the datasets, preprocessing, and experimental setup. Section \ref{results} presents quantitative and qualitative results, including cross-dataset experiments, uncertainty estimation, and ablation studies. Section \ref{discuss} discusses the implications of our findings and their relevance to current research, including potential future directions. Finally, Section \ref{conclusion} summarizes our findings and concludes the study.

\section{Related Works}
\label{related_works}
In this section, we present a review of recent studies on liver tumor segmentation and highlight the use of multi-branch models, attention mechanisms, and multiscale feature extraction approaches, as these have been the recent focus for handling tumor complexity and heterogeneity.

\subsection{Multi-branch and attention-based models}
Automatic segmentation of liver tumors contributes significantly to supporting clinicians in making diagnosis and treatment decisions \cite{rahman2025automatic}. Although Convolutional neural networks (CNN) have been used effectively in medical image segmentation, they struggle to capture long-range pixel dependencies \cite{liu2024s2da}. In contrast, transformer-based models require a substantial number of parameters and involve significant computational expenses \cite{liu2024s2da}. Multibranch and attention-based models have been suggested as a promising alternative to overcome the drawbacks presented by both CNNs and Transformers, as highlighted in recent studies \cite{liu2024s2da, cheng2024eg, wan2025virus, wan2025mhaed, kuang2024adaptive, wang2024sbcnet, dj2024liver, zhang2022decoupled, hu2025msml, oktay2018attention, zhang2021deeprecs, hatamizadeh2021swin}. 

For example, Liu et al. \cite{liu2024s2da} introduced the S2DA-Net, which incorporates a double-branched encoder with a group of multi-head cross-attention aggregation modules to improve accuracy for liver tumors. The model achieved 69.4\% Dice per case and 80.0\% global Dice on LiTS2017 and 3D-IRCADb, respectively. In another work, Cheng et al. \cite{cheng2024eg} proposed an edge-guided segmentation network, EG-UNETR, by synthesizing TransBridge (TB) and a feature semantic enhancement module. The TB utilized multi-head position-aware attention and fusion attention to facilitate the integration of cross-level and multiscale contextual feature fusion. The model scored Dice scores of 84.45\% and 85.01\% on the LiTS2017 and 3D-IRCADb, respectively. Metaheuristic algorithm and multiscale hybrid attention models are also popular in other segmentation tasks \cite{wan2025virus}. For instance, Wan et al. \cite{wan2025mhaed} proposed MHAED-Net that builds an encoder-decoder network with the convolution normalization activation block for efficient segmentation.
Similarly, Kuang et al. \cite{kuang2024adaptive} developed an innovative multiphase channel stacked dual attention module called MCDA, which captures essential semantic information across different phases. The model achieved a Dice score of 79.22\%. On the other hand, Wang et al. \cite{wang2024sbcnet} proposed SBCNet, which incorporates a contextual encoding module to identify tumor variability through the use of a multiscale adaptive kernel, and a boundary enhancement to improve boundary perception by integrating contour learning with the Sobel operator; SBCNet scored a Dice of 82.5\%. In another study, Deepak et al. \cite{dj2024liver} proposed G-Unet, which is capable of incorporating other models such as CNN, residual networks (ResNets), and densely connected convolutional neural networks (DenseNet) into the general U-Net framework. G-Unet reached a global Dice of 72.9\%. In another paper, Zhang et al. \cite{zhang2022decoupled} introduced the Decoupled Pyramid Correlation Network (DPC-Net), which uses a Pyramid Feature Encoder to extract multi-level features and decouples them along spatial and semantic dimensions. The network employed Spatial and Semantic Correlation modules to capture feature correlations, achieving a Dice score of 76.4\% on the LiTS dataset. Finally, Hu et al. \cite{hu2025msml} introduced MSML-AttUNet, a hierarchical attention network that utilizes multilayer dilated convolutions for multiscale feature extraction and expands the receptive field for richer contextual representation; it achieved a Dice score of 87.74\%. 

Oktay et al. \cite{oktay2018attention} introduced Attention U-Net, a spatial attention mechanism to focus on salient features to improve boundary delineation. However, it lacks dynamic modulation based on feature activity, a gap we address using our biologically inspired Synaptic Plasticity Module. TransUNet \cite{zhang2021deeprecs} and Swin UNETR \cite {hatamizadeh2021swin} utilized transformers to model global context, improving long-range dependency capture.
In contrast, Chen et al. \cite{chen2024convnext} proposed ConvNeXt-2U, which integrated two U-Nets with interconnected skip connections and 3D ConvNeXt blocks for segmentation. Nevertheless, these architectures in Euclidean space fail to capture the hierarchical spatial relationships of anatomical structures; this is precisely where our Hyperbolic Convolution Layer excels by embedding features in a non-Euclidean geometric space.

\begin{figure*}[ht!]
    \centering
    \includegraphics[scale=0.24]{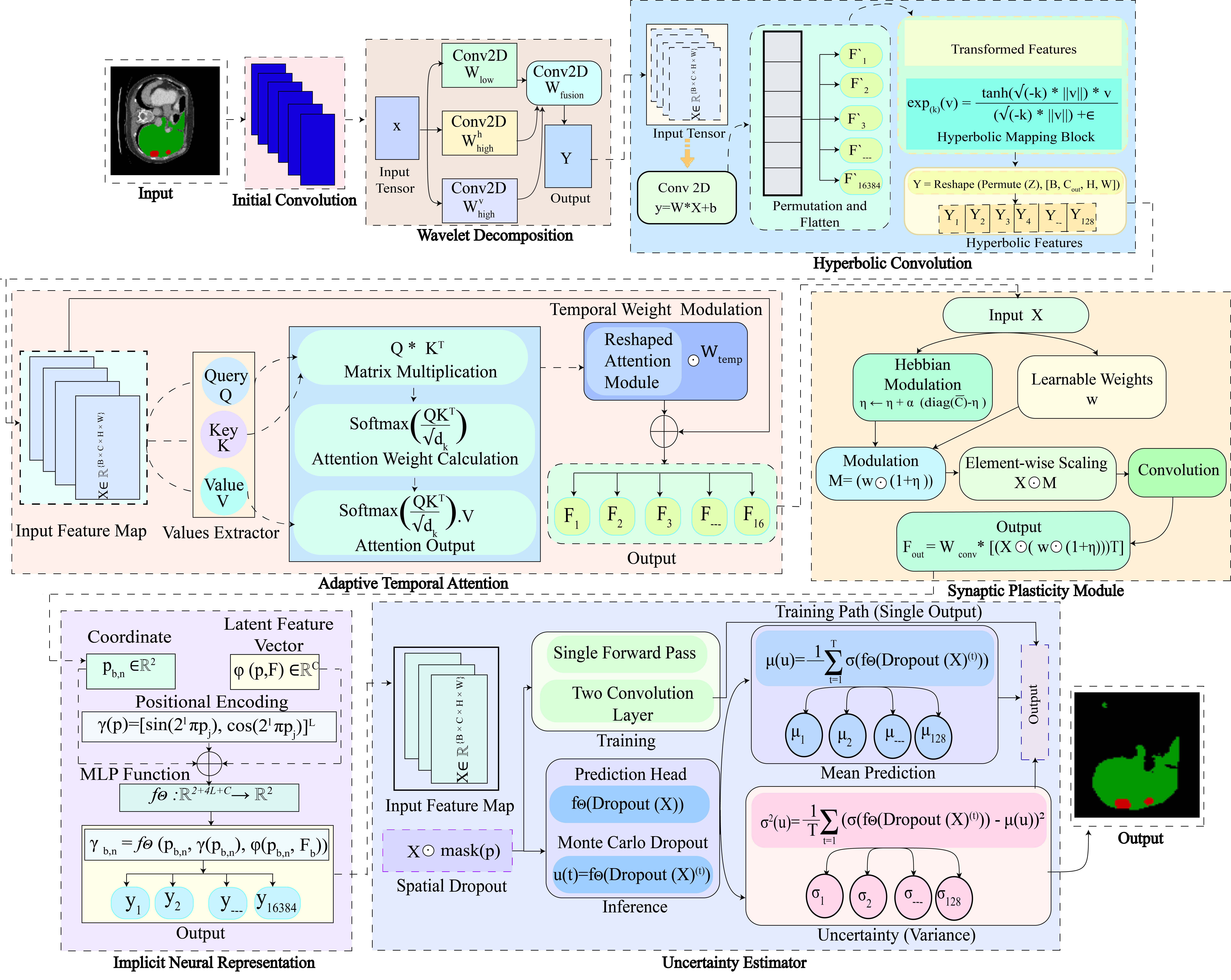}
    \caption{Overview of the proposed HANS-Net architecture that combines hyperbolic convolutions, wavelet decomposition, temporal attention, synaptic plasticity, implicit neural representation, and uncertainty estimator.}
    \label{fig:proposed-system}
\end{figure*}
\subsection{Multi-Phase Fusion}
Recent studies \cite{hu2024trustworthy, cheng2024eg, huang20193d} highlight the effectiveness of models that incorporate temporal information and hierarchical representations in liver tumor segmentation. To begin with, Hu et al. 
\cite{hu2024trustworthy} proposed the TMPLiTS method, a trustworthy multiphase segmentation framework that fuses information from multiple temporal CT scans using a multi-expert mixture scheme, achieving a Dice score of 81.07\%. Similarly, architectures such as EG-UNETR \cite{cheng2024eg} and 3D U²-Net \cite{huang20193d} utilize multiple encoding branches and nested skip connections to integrate multiscale features, allowing both local and global context modeling. Although these models are effective in aggregating spatial information, they often require extensive parameter tuning and remain limited in handling temporal coherence across 3D volumes. In contrast, HANS-Net incorporates a wavelet-inspired decomposition module to facilitate efficient and principled multiscale feature extraction. Beyond spatial scaling, it introduces an Adaptive Temporal Attention mechanism to explicitly model inter-slice dependencies in volumetric scans, an aspect frequently overlooked in slice-based or purely spatial approaches. This dual focus on multiscale structure and temporal dynamics enables HANS-Net to deliver superior segmentation accuracy while preserving anatomical consistency across frames.

\subsection{Geometry and Multiscale Feature Models}
Multiscale and geometry-aware architectures are increasingly explored in recent studies \cite{yang2025multi, you2024contour, yi2022mmf, jha2019resunet++} for their capacity to model complex anatomical variations in liver and tumor segmentation. Yang et al. \cite{yang2025multi} introduced MCFMFNet, a multiscale feature aggregation framework that utilizes multiple convolutional kernels with diverse receptive fields, combined through an element-wise fusion strategy. They also proposed a camouflaged feature mining module to focus attention on tumor regions that are often misclassified as background, achieving Dice scores of 0.8413 and 0.7365 across datasets. Similarly, You et al. \cite{you2024contour}  proposed PGC-Net, which uses a Pyramid Vision Transformer to extract multiscale regional and contour features. These features are refined via a contour-induced parallel graph reasoning mechanism that propagates complementary information through graph convolution and maps it back to the original pixel space. This method achieved Dice scores of 73.63\% and 74.16\% on the LiTS17 and 3DIRCADb datasets, respectively.
Beyond these transformer-based and multiscale CNN approaches, geometry-aware models such as MMF-Net \cite{yi2022mmf} and ResUNet++ \cite{jha2019resunet++} employ pyramid pooling, atrous spatial pyramid modules, and residual refinements to improve contextual understanding across scales. While these designs enhance segmentation performance, they are inherently limited to Euclidean space representations and lack mechanisms to model hierarchical anatomical relationships. Their reliance on stacked convolutional layers can also constrain the ability to capture long-range geometric dependencies, especially in cases of irregular tumor-liver boundaries. In contrast, HANS-Net introduces hyperbolic convolutions into the medical image segmentation domain for the first time, addressing these limitations by enabling compact and hierarchy-aware embeddings. This approach captures the latent anatomical structure more effectively, contributing to improved boundary continuity and segmentation robustness, especially important in anatomically complex regions. Combined with HANS-Net’s wavelet-inspired decomposition and temporal modeling strategies, this geometry-aware design offers a significant advancement over traditional multiscale CNNs and transformer-based methods.

Our proposed approach mitigates several shortcomings found in existing studies by demonstrating novel insights and solutions as summarized in Table \ref{tab:comparison}. Unlike traditional models, our proposed method, HANS-Net, integrates hyperbolic geometry to better capture the hierarchical structure of liver and tumor patterns, which is not addressed by existing methods. Additionally, the wavelet-inspired decomposition module provides multiscale feature extraction without the high computational cost. Furthermore, the inclusion of a biologically inspired synaptic plasticity mechanism allows HANS-Net to adaptively focus on relevant tumor regions. Moreover, the implicit neural representation branch precisely generates boundaries around anatomical regions. Lastly, the introduction of an uncertainty-aware Monte Carlo dropout enables the model to quantify the confidence of prediction. These advances significantly improve the precision, efficiency, and clinical applicability of liver tumor segmentation, effectively addressing critical gaps present in contemporary methodologies.
\section{Methodology}
\label{method}
The overall architecture of HANS-Net is illustrated in Figure \ref{fig:proposed-system}, which outlines the integrated pipeline consisting of six key modules: Wavelet Decomposition, Hyperbolic Convolution, Adaptive Temporal Attention, Synaptic Plasticity Module, Implicit Neural Representation, and Uncertainty Estimation. 

To begin with, the Wavelet-Inspired Decomposition Module splits features into multi-frequency components to preserve both coarse and fine details. Building upon this, the Hyperbolic Convolution Module encodes spatial features into a hyperbolic space to capture hierarchical anatomical relationships. In addition, the Adaptive Temporal Attention module dynamically fuses sequential slices for inter-slice consistency in volumetric data. Furthermore, the Synaptic Plasticity-Enhanced Feature Modulation adaptively strengthens relevant pathways using biologically inspired weight updates. In parallel, the Implicit Neural Representation module enables continuous, high-resolution prediction by modeling outputs as a function of spatial coordinates. Finally, the Uncertainty Estimator, implemented via Monte Carlo Dropout, quantifies prediction confidence to support reliable clinical interpretation. Together, these modules enable HANS-Net to segment liver and tumor structures simultaneously with high accuracy, anatomical boundary fidelity, and robust generalization across datasets.

\begin{figure}[ht!]
    \centering
    \includegraphics[scale = 0.20]{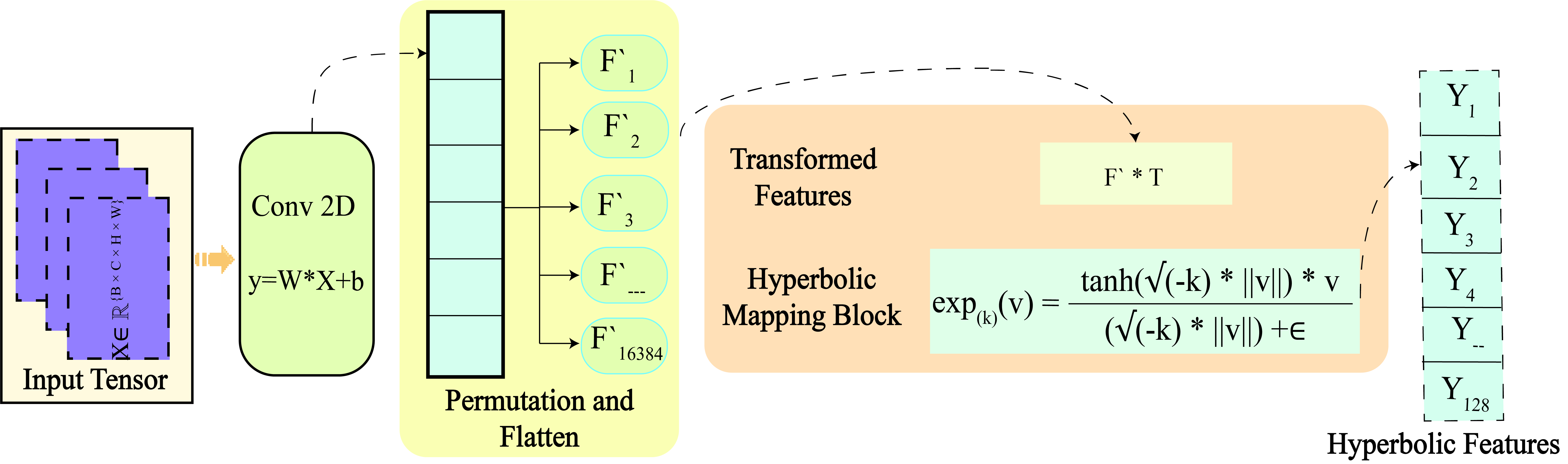}
    \caption{The Hyperbolic Convolution  Module first applies convolution, then flattens and transforms the input before passing it through a hyperbolic mapping block to obtain hyperbolic features.}
    \label{fig:hconv}
\end{figure}
\subsection{Base Model}
Our model processes the input sequentially through a series of specialized modules. It commences with an initial convolution and proceeds with a Wavelet Decomposition Module to extract multi-frequency features. These features are subsequently processed through three hierarchical hyperbolic convolution blocks with pooling to encapsulate structural relationships. The resultant output undergoes refinement via an Adaptive Temporal Attention module and is modulated by a Synaptic Plasticity Module to facilitate enhancement of adaptive features. Concurrently, a parallel Implicit Neural Representation branch was incorporated to capture fine-grained and continuous anatomical boundaries, while an uncertainty estimate quantifies the degree of prediction confidence. 

\subsection{Wavelet Decomposition Module}
We designed the Wavelet Decomposition Module using convolutional layers to perform low-pass and high-pass filtering, followed by a 1$\times$1 convolution to fuse the resulting feature components into a combined representation \cite{liu2021fault}. The overall process can be mathematically formulated in two distinct steps.
First, the input tensor $\mathbf{x} \in \mathbb{R}^{B \times C \times H \times W}$, where $B$ is the batch size, $C$ is the number of input channels, and $H$ and $W$ are the spatial dimensions, is passed through three convolutional filters to extract the low-pass and high-pass features as shown in Equation \eqref{eq:wavelet_filters}:
\begin{align}
    \mathbf{F} = \text{Concat}\Big(
    &~\text{Conv2D}(\mathbf{x}, \mathbf{W}_\text{low}), \nonumber\\
    &~\text{Conv2D}(\mathbf{x}, \mathbf{W}_\text{high}^h), \nonumber\\
    &~\text{Conv2D}(\mathbf{x}, \mathbf{W}_\text{high}^v)
    \Big)
    \label{eq:wavelet_filters}
\end{align}

Next, the concatenated frequency components \( \mathbf{F} \in \mathbb{R}^{B \times C_\text{in} \times H \times W} \) are fused using a \( 1 \times 1 \) convolutional operation, defined in Equation \eqref{eq:wavelet_fusion}:
\begin{equation}
    \mathbf{Y}_{b, c, i, j} = \sum_{k=1}^{C_\text{in}} \sum_{m=1}^{1} \sum_{n=1}^{1} \mathbf{F}_{b, k, i+m-1, j+n-1} \cdot \mathbf{W}^{(c,k)}_{\text{fusion}, m, n} + \mathbf{b}_c
    \label{eq:wavelet_fusion}
\end{equation}

where \( \mathbf{Y} \in \mathbb{R}^{B \times C_\text{out} \times H \times W} \) is the output tensor, \( \mathbf{W}_{\text{fusion}} \in \mathbb{R}^{C_\text{out} \times C_\text{in} \times 1 \times 1} \) is the adaptive convolutional kernel, and \( \mathbf{b} \in \mathbb{R}^{C_\text{out}} \) is an optional bias. Here, \( b \in \{1, \dots, B\} \) indexes the batch, \( c \in \{1, \dots, C_\text{out}\} \) the output channels, \( i, j \in \{1, \dots, H, W\} \) the spatial positions, \( k \in \{1, \dots, C_\text{in}\} \) the input channels, and \( m, n \in \{1\} \) due to the \( 1 \times 1 \) kernel size. This operation performs a weighted channel-wise summation at each spatial location, resulting in the fused output \( \mathbf{Y} \).

The input tensor \( \mathbf{F} \) is formed by concatenating features extracted using frequency-specific filters: \( \mathbf{W}_{\text{low}} \in \mathbb{R}^{C_\text{low} \times K \times K} \) for low-frequency components, \( \mathbf{W}_{\text{high}}^h \in \mathbb{R}^{C_\text{high} \times K \times K} \) for high-frequency horizontal details and \( \mathbf{W}_{\text{high}}^v \in \mathbb{R}^{C_\text{high} \times K \times K} \) for high-frequency vertical features. These spectral components are concatenated to form a rich representation in \( \mathbf{F} \), which is then fused by \( \mathbf{W}_{\text{fusion}} \). This fusion enables the model to integrate multi-frequency information and enhances representational expressiveness. The resulting tensor \( \mathbf{Y} \) serves as the input to the subsequent Hyperbolic Convolution module.

\subsection{Hyperbolic Convolution}
We constructed a HyperbolicConv2d layer to integrate standard convolution with hyperbolic geometry for hierarchical feature learning \cite{bdeir2023fully}. Given an input tensor $\mathbf{x} \in \mathbb{R}^{B \times C_\text{in} \times H \times W}$, where $B$ is the batch size, $C_\text{in}$ is the number of input channels and $H$ and $W$ are the spatial dimensions, a 2D convolution is first applied with an adpatable kernel $\mathbf{W} \in \mathbb{R}^{C_\text{out} \times C_\text{in} \times K \times K}$ to produce the feature map $\mathbf{F}$. Then we flatten the feature map $\mathbf{F}$ along the spatial dimensions and permute to form $\mathbf{F}'$ for transformation in hyperbolic space.
\begin{figure}[!ht]
    \centering
    \includegraphics[scale = 0.185]{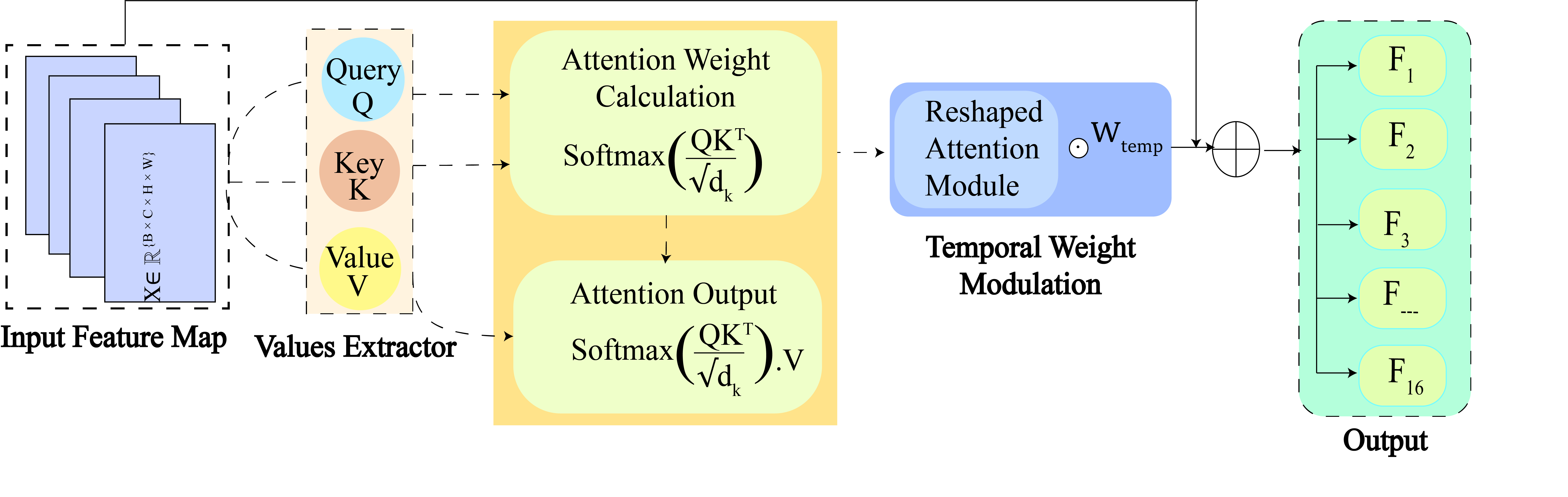}
    \caption{Work flow Temporal Attention Mechanism that processes input features through query-key-value extraction, attention weight calculation, and temporal weight modulation to generate output features}
    \label{fig:attention}
\end{figure}
Next, the permuted feature map $\mathbf{F}'$ is linearly transformed by a able matrix $\mathbf{T} \in \mathbb{R}^{C_\text{out} \times C_\text{out}}$, and the hyperbolic exponential map $\exp_\kappa(\cdot)$ is applied to each feature vector to embed it in the hyperbolic space, as defined in Equation \eqref{eq:expmap}:
\begin{equation}
    \mathbf{Z} = \exp_\kappa(\mathbf{F}' \mathbf{T}), 
    \quad 
    \exp_\kappa(\mathbf{v}) = 
    \frac{
        \tanh\!\big( \sqrt{-\kappa} \, \| \mathbf{v} \| \big)
    }{
        \sqrt{-\kappa} \, \| \mathbf{v} \| + \epsilon
    } 
    \mathbf{v}
    \label{eq:expmap}
\end{equation}
where $\kappa$ is the curvature parameter of the hyperbolic space, $\| \mathbf{v} \|$ refers to the Euclidean norm of the vector $\mathbf{v}$, and $\epsilon$ is a small constant for numerical stability. Finally, the hyperbolic transformed features $\mathbf{Z}$ are permuted and reshaped to recover the original spatial dimensions, producing the final output $\mathbf{Y} \in \mathbb{R}^{B \times C_\text{out} \times H \times W}$, as described in Equation \eqref{eq:reshape}:
\begin{equation}
    \mathbf{Y} = \text{Reshape}(\text{Permute}(\mathbf{Z}), [B, C_\text{out}, H, W])
    \label{eq:reshape}
\end{equation}
As illustrated in Figure \ref{fig:hconv}, the Hyperbolic Convolution Module projects feature maps into a hyperbolic space to better capture the latent anatomical hierarchy. This embedding enables compact representations of tree-like tumor boundaries and nested liver structures, which are challenging for Euclidean CNNs. The module output
$\mathbf{Y}$ serves directly as the input for the subsequent Adaptive Temporal Attention stage. 

\subsection{Adaptive Temporal Attention}
In volumetric data, the slice thickness and spacing can vary across scanners and acquisition protocols. In our study, all scans are first resampled following area-based resampling and nearest-neighbor interpolation during preprocessing for uniformity; and to handle inter-slice dependencies across CT volumes, we introduce an Adaptive Temporal Attention (ATA) mechanism, shown in Figure \ref{fig:attention}. This ATA dynamically weighs neighboring slices based on learned temporal importance, allowing the network to enhance spatial continuity across volumes while maintaining computational efficiency \cite{shi2020spatial}. Unlike fully 3D convolutional approaches that process entire volumes simultaneously at high computational cost, our ATA mechanism operates on 2D slices while explicitly modeling inter-slice relationships through learned attention weights, thereby preserving volumetric continuity without sacrificing efficiency. The temporal attention component effectively reconstructs inter-slice dependencies by attending to contextually relevant neighboring slices during feature extraction. Equation \eqref{eq:ATA} formulates the attention-enhanced output: 
\begin{equation}
\mathbf{F}_{\text{out}} = \mathbf{X} + \left(\mathbf{W}_{\text{temp}} \odot \text{reshape}\left(\text{softmax}\left(\frac{\mathbf{Q}\mathbf{K}^{\top}}{\sqrt{d_k}}\right) \cdot \mathbf{V}\right)\right)
\label{eq:ATA}
\end{equation}
where $\mathbf{X} \in \mathbb{R}^{B \times C \times H \times W}$ is the input feature map, $B$ is the batch size, $C$ is the number of channels, and $H \times W$ represents the spatial resolution. The query $\mathbf{Q}$, key $\mathbf{K}$, and value $\mathbf{V}$ tensors are derived from $\mathbf{X}$ using optimized $1 \times 1$ convolutions and reshaped to $\mathbb{R}^{B \times N \times d_k}$, where $N$ is the dot product of $H \cdot W$, and $d_k = C/8$ is the reduced embedding dimension. A scaled dot product attention mechanism computes spatial attention weights via $\text{softmax}(\mathbf{Q}\mathbf{K}^{\top}/\sqrt{d_k})$, which are then applied to $\mathbf{V}$ to generate an attention-weighted representation. This output is reshaped to the original spatial format and further modulated by a trained temporal weight vector $\mathbf{W}_{\text{temp}} \in \mathbb{R}^{B \times 1 \times 1 \times 1}$, derived from the temporal context using a multilayer perceptron. Thus, instead of assuming uniform temporal spacing, ATA dynamically adjusts the contribution of neighboring slices according to their feature correlations. This allows the attention map to dynamically adjust for non-uniform slice intervals, including irregular thickness, by modulating contextual influence according to observed spatial correlations rather than fixed slice indices. Furthermore, by extracting slices from multiple anatomical views, the ATA mechanism captures complementary spatial information from orthogonal directions, effectively approximating the comprehensive context of full 3D analysis. This multi-view approach with adaptive temporal weighting maintains robust inter-slice consistency while offering significant computational efficiency and reduced memory requirements.

The result is then fused with the original input $\mathbf{X}$ through a residual connection, producing $\mathbf{F}_{\text{out}}$, which is enriched in both spatial and temporal dimensions. The output generated by this module is then fed into the Synaptic Plasticity unit.

\subsection{Synaptic Plasticity Mechanism}
We designed a synaptic plasticity mechanism inspired by Hebbian learning to enable biologically driven adaptive feature modulation within the framework \cite{wang2021computational}. This mechanism is based on the fundamental principle of activity-dependent synaptic strengthening observed in neural systems, where connections between co-activated neurons are reinforced over time. While our implementation represents a computational approximation optimized for gradient-based learning rather than a direct model of neurophysiological processes, it incorporates the core concept of correlation-based weight adjustment into a differentiable framework that enhances feature adaptivity in medical image segmentation \cite{hansel2024neural}. Given an input feature tensor $\mathbf{X}$, the module first computes a channel-wise activity vector for each sample. Specifically, for each sample $b$ in batch $B$, we obtain a channel activity vector $\mathbf{a}_b \in \mathbb{R}^{C}$ by globally averaging the spatial responses of the feature tensor $b^{th}$. Next, for each sample, we computed a correlation matrix $\mathbf{C}_b \in \mathbb{R}^{C \times C}$ as the outer product of the activity vector $a_b$ with itself. 

The batch-averaged correlation matrix $\mathbf{C} \in \mathbb{R}^{C \times C}$ is then obtained by averaging all $\mathbf{C}_b$ over the batch dimension. The diagonal elements of $\mathbf{C}$, which represent the self-correlation of each channel, are used to update the synaptic trace vector $\boldsymbol{\eta'} \in \mathbb{R}^{C}$ following the Hebbian update rule, as shown in Equation \eqref{eq:hebb_trace}:
\begin{equation}
    \boldsymbol{\eta}' \leftarrow \boldsymbol{\eta} + \alpha \, \text{diag}(\mathbf{C}) - \boldsymbol{\eta}
    \label{eq:hebb_trace}
\end{equation}
where $\boldsymbol{\eta}$ is a trace vector; $\alpha$ is the plasticity learning rate that regulates the update process; and $\text{diag}(\mathbf{C})$ is the vector formed by the diagonal entries of $\mathbf{C}$. This update mechanism mirrors the strengthening of synaptic weights in response to correlated neural activity, providing an inductive bias that encourages the network to adaptively emphasize informative feature channels during training. The updated trace vector $\boldsymbol{\eta'}$ modulates a set of learnable synaptic scaling factors $\boldsymbol{\omega} \in \mathbb{R}^{C}$, producing an adaptive channel-wise gain. This gain adjusts the input tensor $\mathbf{X}$ element-wise. The modulated features are then linearly transformed by a learnable transformation matrix $\mathbf{T} \in \mathbb{R}^{C \times C}$ to capture cross-channel interactions before convolution. Finally, as shown in Equation \eqref{eq:hebb_output}, a convolution with kernel weights $\mathbf{W}_\text{conv} \in \mathbb{R}^{C_\text{out} \times C \times K \times K}$ refines the transformed features to produce the final output tensor $\mathbf{F}_\text{out}$:
\begin{equation}
    \mathbf{F}_\text{out} = \mathbf{W}_\text{conv} * \big[ (\mathbf{X} \odot (\boldsymbol{\omega} \odot (1 + \boldsymbol{\eta})) ) \mathbf{T} \big]
    \label{eq:hebb_output}
\end{equation}
where $*$ denotes the convolution operation, $\odot$ represents element-wise multiplication, and $\mathbf{T}$ represents the channel-wise modulated features. This formulation ensures that the output feature map $\mathbf{F}_\text{out}$  captures not only the information coming from the input, but also the adaptive changes from synaptic plasticity, making the network more realistic and better adapted to learn. The output feature map $\mathbf{F}_\text{out}$ from this stage now acts as the input to the Implicit Neural Representation module.

\subsection{Implicit Neural Representation }
We developed a continuous coordinate-based neural representation module to support high-resolution segmentation predictions \cite{zhang2022implicit}. For each normalized spatial coordinate $\mathbf{p}_{b,n} \in \mathbb{R}^2$ from the $b$th image in the batch, where values lie within the range $[-1, 1]^2$, we apply a positional encoding function $\gamma(\cdot)$ to effectively capture spatial information across multiple scales. This function encodes each coordinate dimension using sine and cosine functions with progressively increasing frequencies, as described in Equation \eqref{eq:posenc}:
\begin{equation}
\gamma(\mathbf{p}) = 
\big[ 
\sin(2^l \pi p_j), \;
\cos(2^l \pi p_j)
\big]
\quad \text{for}~ j = 1, 2;~ l = 0, \ldots, L-1
\label{eq:posenc}
\end{equation}

In parallel, for each coordinate $\mathbf{p}$, a latent feature vector $\varphi(\mathbf{p}, \mathbf{F}) \in \mathbb{R}^C$ is extracted by bilinearly sampling from the global feature map $\mathbf{F}_b \in \mathbb{R}^{C \times H \times W}$ using grid sampling. The multi-layer perceptron $f_\theta$ maps the concatenation of the coordinate, its positional encoding, and the sampled feature vector to a 2D logit, as specified in Equation \eqref{eq:mlpmap}:
\begin{equation}
f_\theta : \mathbb{R}^{2 + 4L + C} \longrightarrow \mathbb{R}^2
\label{eq:mlpmap}
\end{equation}

\begin{figure}[ht!]
    \centering
    \includegraphics[scale=0.33]{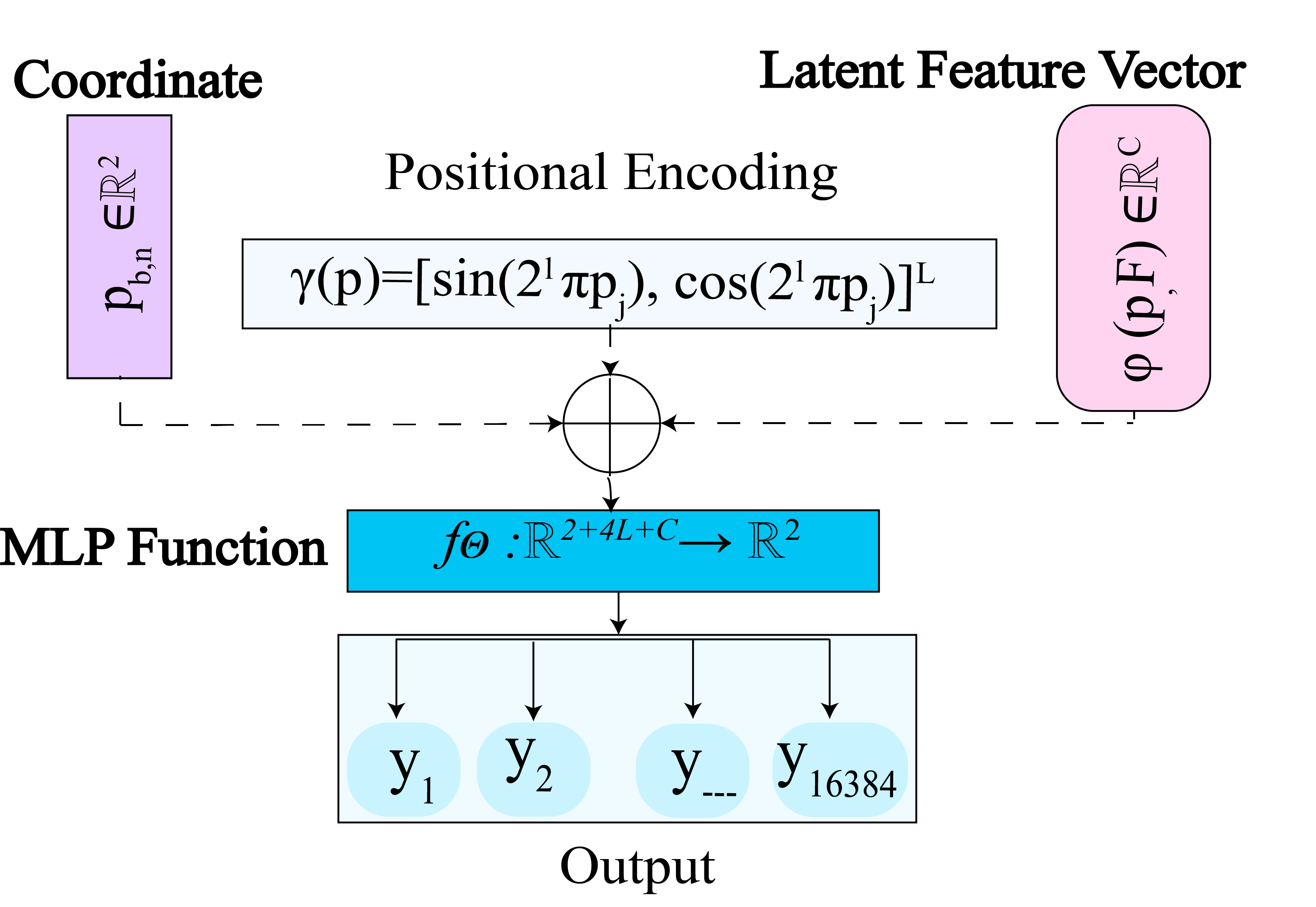}
    \caption{The process of implicit neural representation, comprised of encoding coordinates with positional encoding, combining them with a latent feature vector, and passing the result through an MLP function to generate the output.}
    \label{fig:Neural}
\end{figure}

Finally, the network output for the logit at each spatial coordinate is computed using this MLP, as described in Equation \eqref{eq:mlppredict}:
\begin{equation}
\mathbf{y}_{b,n} = f_\theta 
\big( 
\mathbf{p}_{b,n}, ~ 
\gamma(\mathbf{p}_{b,n}), ~ 
\varphi(\mathbf{p}_{b,n}, \mathbf{F}_b)
\big)
\label{eq:mlppredict}
\end{equation}
This design enables precise, continuous spatial prediction without relying on a fixed-resolution grid. The resulting tensor produced by this module is subsequently utilized as the input for the Uncertainty Estimator. Figure \ref{fig:Neural} demonstrates the working process of Implicit Neural Representation.
\subsection{Uncertainty Estimator}
\begin{figure}[ht!]
    \centering
    \includegraphics[scale=0.15]{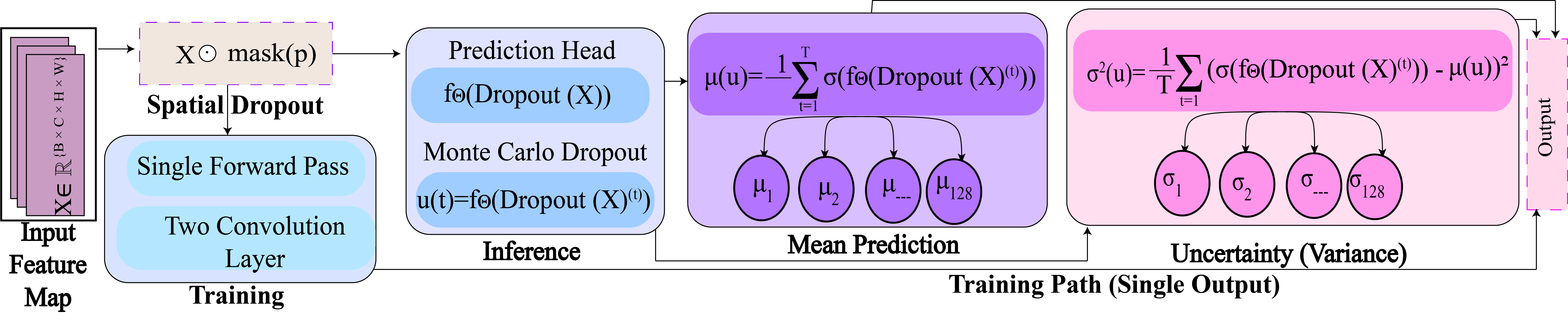}
    \caption{The process of uncertainty estimator using Monte Carlo Dropout, where the input feature map undergoes spatial dropout, followed by inference, mean prediction, and uncertainty estimator}
    \label{fig:estimator}
\end{figure}

We have constructed an uncertainty-aware prediction module (see Figure \ref{fig:estimator}) based on Monte Carlo dropout to estimate both predictive output and its associated confidence \cite{tang2022prediction}. Given an input feature map $\mathbf{X}$, we first apply a spatial dropout with probability $p$ to simulate stochastic behavior at the test time. During training, a single forward pass is performed applying dropout followed by a lightweight prediction head $f_\theta(\cdot)$, consisting of two convolutional layers and a sigmoid activation, which outputs a confidence map in pixels. During inference, to quantify epistemic uncertainty, we perform stochastic forward passes $T$ by applying dropout independently at each iteration, these passes are indexed by a small variable $t \in \{1, 2, \ldots, T\}$, where each $u^{(t)}$ in $f_\theta(\text{Dropout}(X)^{(t)})$ denotes the model prediction from the $t^{\text{th}}$ forward pass with a distinct dropout mask. 
The mean prediction $\mu(\mathbf{u})$ is obtained by averaging these outputs, while the uncertainty is quantified by the variance $\sigma^2(\mathbf{u})$, calculated as the pixel variance across the $T$ samples. The formulations are provided in Equations \eqref{eq:youu} and \eqref{eq:you}:
\begin{align}
\mu(\mathbf{u}) &= \frac{1}{T} \sum_{t=1}^{T} \sigma\left(f_\theta\left(\text{Dropout}(\mathbf{X})^{(t)}\right)\right) \label{eq:youu}\\
\sigma^2(\mathbf{u}) &= \frac{1}{T} \sum_{t=1}^{T} \left(\sigma\left(f_\theta\left(\text{Dropout}(\mathbf{X})^{(t)}\right)\right) - \mu(\mathbf{u})\right)^2
\label{eq:you}
\end{align}
This approach enables the model to express uncertainty in a principled, data-driven manner without requiring architectural changes or additional supervision. 

\section{Experimental Details}
\label{experiment}
In this section, we elaborate on the dataset preparation tasks along with the model's preprocessing and implementation setup.
\subsection{Data Preparation}
\subsubsection{Datasets}
In our study, we used two public datasets, including LiTS and AMOS 2022, to evaluate our model's performance and generalization capability. It is worth noting that since the official LiTS test set (70 samples) was not publicly available, we limited our analysis to the 131 publicly released training cases to ensure full reproducibility for all researchers. We randomly split these available 131 samples into an 80:20 ratio, resulting in 105 cases for training and 26 cases for testing. To demonstrate the cross-dataset generalization and modality robustness of our model, we also evaluated our model on the AMOS 2022 dataset, which contains both CT and MRI scans with 15 annotated abdominal organs. Although the AMOS 2022 dataset has a larger number of samples, we randomly selected 70 samples (unseen by the model), containing both CT and MRI scans, for cross-dataset validation to assess the model’s robustness across different imaging modalities and anatomical variations.

\noindent\textbf{Liver Tumor Segmentation Challenge (LiTS).} The LiTS dataset \cite{bilic2023liver} comprises 131 (publicly available) volumetric abdominal CT scans, each with a pixel resolution of $512 \times 512$ and varying voxel spacings, with in-plane resolutions between 0.60 and 0.98 mm and slice thicknesses ranging from 0.45 to 5.0 mm. It includes a variety of liver-related abnormalities, such as tumors, metastases, and cysts. Tumor volumes vary from 38 mm\textsuperscript{3} to 1231 mm\textsuperscript{3}, and the number of tumors per scan ranges from 0 to 12. From these scans, we extract a total of 15,654 2D tumor-containing slices from all three anatomical views (4,108 axial, 5,730 coronal, and 5,816 sagittal). These selected slices are then divided into training (10,957), validation (2,348), and test (2,349) sets, ensuring that both liver parenchyma and tumor regions are present for effective segmentation learning.

\noindent\textbf{AMOS 2022.} The AMOS dataset \cite{ji2022amos} is a large-scale abdominal multi-organ benchmark containing 500 CT and 100 MRI scans collected from multi-center, multi-scanner clinical environments. Each scan provides voxel-level annotations for 15 abdominal organs, including the spleen, kidneys, gallbladder, esophagus, liver, stomach, aorta, postcava, pancreas, adrenal glands, duodenum, bladder, and prostate. The dataset features multi-modality, multi-phase, and multi-disease cases, capturing a broad spectrum of imaging conditions and pathological variations.
\subsubsection{Data Preprocessing}
\label{data_proc}
We preprocess the 3D medical images to standardize their format and intensity ranges. Area-based resampling is applied to CT volumes and nearest-neighbor interpolation to segmentation masks to preserve anatomical boundaries, while retaining only slices with visible tumors, including subtle lesions. Windowing is applied by clipping each voxel intensity to the range \([-200, 400]\) and linearly rescaling it to \([0, 1]\), enhancing contrast around soft tissues. Intensity normalization is applied after windowing by dividing the clipped value range into equal intervals across all slices. This preprocessing is performed across axial, coronal, and sagittal views, with each 2D slice extracted from the original 3D volume at a resolution of \(512 \times 512\). Our framework processes these slices at a resized in-plane resolution of \(128 \times 128\) due to computational resource constraints. This resizing is applied uniformly to all slices to maintain depth and ensure uniform spatial dimensions \cite{salvi2021impact}.

\subsection{Implementation Setup}
This research is conducted using an AMD Ryzen 5 5600X 6-core Central Processing Unit (CPU) and 16 GB of RAM for all experiments. For graphical processing, a ZOTAC GAMING GeForce RTX 3060 Twin Edge OC with 12 GB of video RAM (VRAM) is utilized, while Jupyter Notebook version 7.0.8 serves as the integrated development environment (IDE).

\section{Results}
\label{results}

\subsection{Performance Assessment}

\noindent\textbf{Segmentation performance of the proposed model.} To evaluate the segmentation performance of our proposed HANS-Net framework, we conducted comprehensive experiments on the LiTS dataset, focusing on accurate liver and tumor delineation. The architecture integrates hyperbolic convolutions, wavelet-inspired decomposition, synaptic plasticity mechanisms, and implicit neural representations to capture both global organ structure and local tumor details within a continuous and geometrically informed space. All CT volumes were resampled to a uniform resolution of 128~$\times$~128 pixels, and the model was trained for 100 epochs using the Adam optimizer with a learning rate of 0.001 and a batch size of 16. 
\begin{table}[!ht]
\centering
\caption{Segmentation performance of the proposed HANS-Net model on the LiTS dataset for liver and tumor structures.}
\begin{tabular}{lcccc}
\midrule
\textbf{Organ} & \textbf{Dice (\%)} & \textbf{IoU (\%)} & \textbf{ASSD (mm)} & \textbf{VOE (\%)} \\
\midrule
Liver & 96.67 & 93.79 & 0.54 & 6.21 \\
Tumor & 89.84 & 82.39 & 0.89 & 17.61 \\
\rowcolor{gray!20}Mean & 93.26 & 88.09 & 0.72 & 11.91 \\
\midrule
\end{tabular}
\label{quantitive_seg}
\end{table}
As reported in Table~\ref{quantitive_seg}, HANS-Net achieves a Dice score of 96.67\% for liver and 89.84\% for tumors, demonstrating high segmentation precision across both large organ regions and smaller tumor structures. The corresponding IoU values of 93.79\% (liver) and 82.39\% (tumor) confirm strong spatial overlap between predictions and ground truth. Furthermore, in terms of boundary alignment, the model attains a low Average Symmetric Surface Distance (ASSD) of 0.54 mm for the liver and 0.89 mm for tumors, while the Volume Overlap Error (VOE) remains limited to 6.21\% and 17.61\%, respectively. The observed performance gap between liver and tumor segmentation reflects the inherent difficulty of tumor delineation due to substantial heterogeneity in size, shape, contrast, and diffuse boundaries. The synaptic plasticity module and implicit neural representation branch specifically address these challenges by adaptively refining tumor-relevant features and modeling continuous boundaries. These results validate that HANS-Net effectively integrates geometric and biologically inspired components to produce accurate and anatomically consistent liver tumor segmentation.

\begin{figure*}[ht!]
\centering
\includegraphics[scale=0.063]{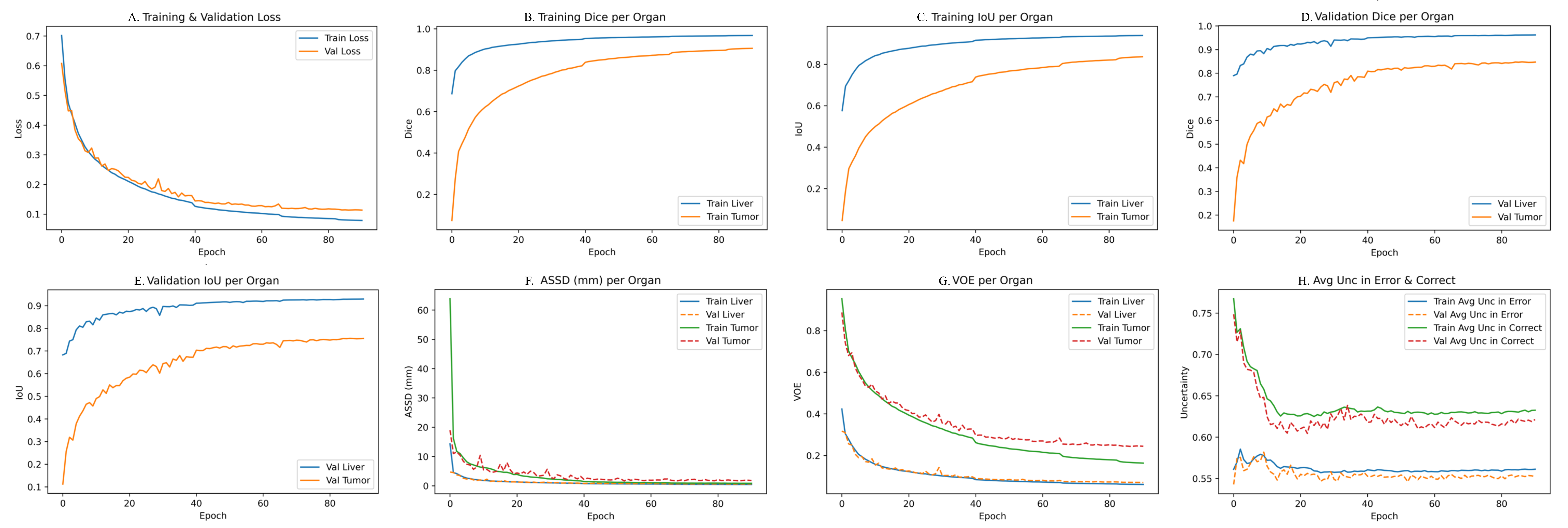}
\caption{Accuracy and loss curves of our proposed HANS-Net model.}
\label{curves}
\end{figure*}

\noindent\textbf{Training Dynamics and Convergence.} The training and validation curves for our proposed HANS-Net model demonstrate effective convergence and stable performance in liver and tumor segmentation tasks. As shown in Figure \ref{curves}A, both training and validation losses decreased steadily from initial values around 0.7 to approximately 0.1, with training loss converging slightly faster than validation loss, indicating good model generalization without significant overfitting. The progression of the Dice score reveals distinct learning patterns between organ types. Training Dice scores for liver segmentation in Figure \ref{curves}B rapidly increased from 10\% to more than 90\% within the first 20 epochs, demonstrating the model's ability to quickly capture large anatomical structures. 

In contrast, tumor Dice scores showed a more gradual improvement, reaching approximately 80\% by epoch 80, reflecting the inherent difficulty of segmenting smaller, more heterogeneous tumor regions. Validation Dice scores in Figure \ref{curves}D followed similar trends, with liver achieving over 90\% and tumor reaching approximately 80\%, confirming robust generalization performance. IoU metrics in Figure \ref{curves}C and \ref{curves}E mirrored the Dice score patterns, with liver IoU rapidly increasing to 90\% during training and tumor IoU gradually improving to 75\%. The faster convergence of liver metrics compared to tumor metrics highlights the model's hierarchical learning approach, where larger anatomical structures are learned before fine-grained tumor boundaries. The ASSD values in Figure \ref{curves}F decreased to minimal levels below 5mm for both organs, indicating a precise boundary delimitation. The VOE metrics in Figure \ref{curves}G decreased substantially during early training, stabilizing around 0.2 for the liver and 0.3 for the tumor, while the analysis in Figure \ref{curves}H showed decreasing uncertainty levels for both correct and incorrect predictions, and the model became more confident in its predictions as training progressed. These comprehensive training dynamics validate that our HANS-Net framework effectively integrates hyperbolic convolutions, wavelet decomposition, and synaptic plasticity mechanisms to achieve robust liver tumor segmentation with distinct learning trajectories optimized for different anatomical structures.

\subsection{Comparison With State-of-the-Art Methods}
\label{compSOTA}
We compared the proposed HANS-Net model with several SOTA liver and tumor segmentation methods, including U-Net \cite{ronneberger2015u}, U-Net++ \cite{zhou2018unet++}, Attention U-Net \cite{oktay2018attention}, H-DenseUNet \cite{li2018h}, TransUNet \cite{chen2021transunet}, and nnU-Net \cite{isensee2021nnu}. To ensure fair comparison with architectures originally designed for 3D or hybrid 2D/3D processing, we adapted them to our slice-based evaluation by processing CT slices individually through their 2D encoder-decoder pathways while retaining original components such as attention gates and dense blocks. Attention U-Net's spatial attention was applied at each decoder level, and H-DenseUNet's hybrid dense blocks were implemented in 2D convolutional layers with consistent hyperparameters. All baselines used the same preprocessing, multi-view slice extraction, Adam optimizer with a learning rate of 1e-4, batch size of 16, and training for 100 epochs, ensuring a consistent evaluation framework across all models. The proposed HANS-Net framework consistently surpasses all SOTA methods in most metrics. Although TransUNet \cite{chen2021transunet} achieved a marginally lower ASSD of 0.51 mm for liver, where H-DenseUNet \cite{li2018h} recorded a slightly lower VOE of 17.54\% for tumors. The nnU-Net \cite{isensee2021nnu}, on the other hand, showed competitive liver segmentation performance and achieved a second-highest Dice score of 96.20\% in the liver segmentation task \cite{bilic2023liver}. Compared to the existing models, our model achieved the highest Dice scores of 96.67\% for liver and 89.84\% for tumor, which is +0.27\%, and +2.92\% Dice score improvement compared to nnU-Net and TransUNet, respectively. The model also recorded the highest IoU scores of 93.79\% for liver and 82.39\% for tumor, and the lowest average ASSD of 0.72 mm and VOE of 11.91\%.

\begin{table}[!ht]
\centering
\footnotesize
\caption{Comparative performance of HANS-Net and SOTA models on the LiTS dataset for liver and tumor segmentation. Metrics include Dice (\%), IoU (\%), ASSD, and VOE (\%). The best and second-best results for each dataset are \textbf{bolded} and \underline{underlined}, respectively.}
\label{comparative_performance_liver}
\begin{scriptsize}
\begin{tabular}{llcccc}
\toprule
\textbf{Model} & \textbf{Organ} & \textbf{Dice $\uparrow$} & \textbf{IoU $\uparrow$} & \textbf{ASSD(mm) $\downarrow$} & \textbf{VOE $\downarrow$} \\
\midrule
\multirow{3}{*}{U-Net \cite{ronneberger2015u}}  
& Liver & 92.18 & 86.21 & 1.32 & 13.79 \\
& Tumor & 82.36 & 74.52 & 1.76 & 25.48 \\
& Mean  & 87.27 & 80.37 & 1.54 & 19.64 \\
\midrule
\multirow{3}{*}{U-Net++ \cite{zhou2018unet++}}  
& Liver & 93.01 & 87.63 & 1.12 & 12.37 \\
& Tumor & 83.49 & 76.05 & 1.62 & 23.95 \\
& Mean  & 88.25 & 81.84 & 1.37 & 18.16 \\
\midrule
\multirow{3}{*}{Attention U-Net \cite{oktay2018attention}}  
& Liver & 94.07 & 89.24 & 0.98 & 10.76 \\
& Tumor & 85.34 & 78.18 & 1.42 & 21.82 \\
& Mean  & 89.71 & 83.71 & 1.20 & 16.29 \\
\midrule
\multirow{3}{*}{H-DenseUNet \cite{li2018h}}  
& Liver & 93.63 & 88.72 & 1.04 & 11.28 \\
& Tumor & 84.89 & 77.03 & 1.51 & \textbf{17.54} \\
& Mean  & 89.26 & 82.88 & 1.28 & 17.41 \\
\midrule
\multirow{3}{*}{TransUNet \cite{chen2021transunet}}  
& Liver & 94.82 & \underline{90.13} & \textbf{0.51} & \underline{9.87} \\
& Tumor & \underline{86.92} & \underline{79.88} & 1.33 & 20.12 \\
& Mean  & 90.87 & 85.01 & 0.92 & 15.00 \\
\midrule
\multirow{3}{*}{nnU-Net} \cite{isensee2021nnu}  
&Liver & \underline{96.20} & 89.36 & 2.56 & 10.44 \\
& Tumor & 73.90 &  78.93 & \underline{0.90} & 19.96 \\
& Mean & 85.05 &  85.22 & 1.73 & 15.20 \\
\midrule
\multirow{3}{*}{\textbf{HANS-Net (ours)}}  
& Liver & \textbf{96.67} & \textbf{93.79} & \underline{0.54} & \textbf{6.21} \\
& Tumor & \textbf{89.84} & \textbf{82.39} & \textbf{0.89} & \underline{17.61} \\
\rowcolor{gray!20} & Mean  & 93.26 & 88.09 & 0.72 & 11.91 \\
\bottomrule
\end{tabular}
\end{scriptsize}
\end{table}

\begin{figure}[ht!]
\centering
\includegraphics[scale=0.033]{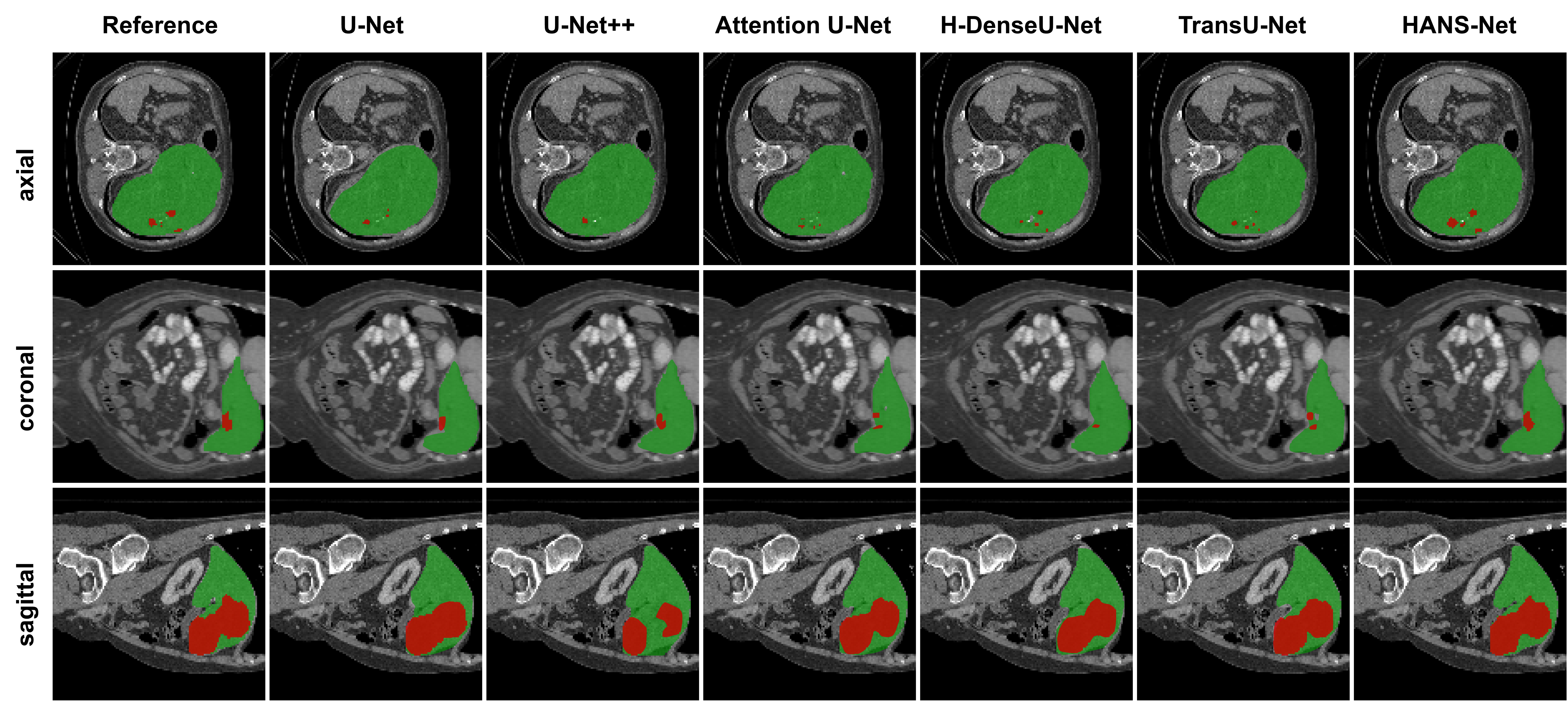}
\caption{Comparative performance of the proposed HANS-Net model and baseline methods (U-Net, U-Net++, Attention U-Net, H-DenseUNet, and TransUNet) for liver and tumor segmentation on the LiTS dataset.}
\label{model_pred}
\end{figure}

Furthermore, the mean segmentation quality between both organs is the most consistent and accurate, confirming the superiority of HANS-Net in anatomical boundary preservation and tumor detection accuracy. Table \ref{comparative_performance_liver} presents detailed comparative performance in Dice, IoU, ASSD, and VOE, demonstrating the superiority of our HANS-Net framework in accurate, robust, and anatomically consistent liver and tumor segmentation. Figure \ref{model_pred} visualizes the superior segmentation capabilities of our approach compared to the SOTA methods tested.

\subsection{Comparison With Existing Literature}
Our proposed method, HANS-Net, demonstrates superiority over existing models by achieving the highest Dice score of 93.26\% on the LiTS dataset. This outcome proves HANS-Net's enhanced capability in the segmentation of liver tumors compared to recent SOTA methods. 

Previous models, such as MSML-AttUNet \cite{hu2025msml}, attained performances of 87.74\% on combined datasets like LiTS and 3DIRCADb. Although models like EG-UNETR \cite{cheng2024eg} exhibited commendable performance at 85.01\% and S2DA-Net \cite{liu2024s2da} demonstrated moderate results with 80.0\%, HANS-Net presents a marked improvement. Models such as MCDA \cite{kuang2024adaptive} and G-Unet \cite{dj2024liver} recorded lower scores, suggesting constraints in generalization or segmentation precision. Even methodologies employing private datasets, such as MCFMFNet \cite{yang2025multi}, did not exceed the 90\% threshold. The comparative analysis highlights the strength and practical effectiveness of HANS-Net, showing that it has a distinct advantage over the existing SOTA models. Table \ref{tab:comparison} provides a detailed comparison between the proposed approach and the SOTA techniques.

\begin{table}[ht!]
\centering
\small
\caption{Comparison of different models and their Dice scores on various datasets. The best and second-best results for each dataset are \textbf{bolded} and \underline{underlined}, respectively.}
\label{tab:comparison}
\begin{tabular}{cllc}
\midrule
\textbf{Ref.} & \textbf{Model} & \textbf{Dataset} & \textbf{Dice} \\
\midrule
\cite{wang2024sbcnet} & SBCNet & LiTS & 82.50\% \\
\cite{yang2025multi} & MCFMFNet & LiTS & 84.13\% \\
\cite{liu2024s2da}   & S2DA-Net & LiTS & 80.00\% \\
\cite{cheng2024eg} & EG-UNETR & LiTS & \underline{84.45\%} \\
\cite{dj2024liver} & G-Unet & LiTS & 72.90\% \\
\cite{you2024contour} & PGC-Net & LiTS & 73.63\% \\
\cite{zhang2022decoupled} & DPC-Net & LiTS &76.40\% \\
\cite{cheng2024eg} & EG-UNETR & 3D-IRCADb & 85.01\% \\
\cite{you2024contour} & PGC-Net & 3D-IRCADb & 74.16\% \\
\cite{yang2025multi} & MCFMFNet & Private & 73.65\% \\
\cite{kuang2024adaptive} & MCDA & Private & 79.22\% \\
\cite{hu2024trustworthy} & TMPLITS & Private & 81.07\% \\
\midrule
Ours & HANS-Net & LiTS & \textbf{93.26}\% \\
     &       & AMOS 2022 & \underline{ 85.09\%} \\
\midrule
\end{tabular}
\end{table}

\subsection{Cross-Dataset Evaluation} 

\begin{table*}[!ht]
\centering
\caption{Segmentation performance of HANS-Net on the AMOS 2022 dataset across 15 abdominal organs. The abbreviated organs are: Right Kidney (R. Kid.); Left Kidney (L. Kid.); Stomach (Stom.); Pancreas (Panc.); Right Adrenal Gland (R. Adr.); Left Adrenal Gland (L. Adr.); Duodenum (Duod.); Urinary Bladder (Blad.); Prostate (Prost.).}
\label{tab:quantitive_seg_amos}
\fontsize{6.5pt}{8pt}\selectfont

\begin{tabular}{lcccccccccccccccc}
\toprule
\textbf{Metric} & \textbf{Spleen} & \textbf{R. Kid.} & \textbf{L. Kid.} & \textbf{Gallbladder} & \textbf{Esophagus} & \textbf{Liver} & \textbf{Stom.} & \textbf{Aorta} & \textbf{Postcava} & \textbf{Panc.} & \textbf{R. Adr.} & \textbf{L. Adr.} & \textbf{Duod.} & \textbf{Blad.} & \textbf{Prost.} & \textbf{Mean} \\
\midrule
Dice (\%)  & 81.37 & 83.36 & 80.85 & 88.56 & 81.51 & 94.17 & 85.52 & 80.96 & 94.13 & 86.23 & 87.65 & 87.90 & 80.72 & 83.53 & 79.89 & \textbf{85.09} \\
IoU (\%)  & 75.93 & 78.84 & 73.16 & 75.58 & 75.06 & 80.12 & 74.42 & 69.83 & 88.94 & 79.48 & 84.32 & 75.24 & 71.05 & 77.90 & 70.08 & \textbf{76.66} \\
ASSD (mm) & 20.63 & 26.31 & 20.37 & 15.96 & 24.94 & 8.35 & 16.72 & 25.89 & 9.65 & 18.88 & 17.75 & 13.53 & 21.39 & 19.44 & 32.49 & \textbf{19.49} \\
VOE (\%)  & 24.07 & 21.16 & 26.84 & 24.42 & 24.94 & 19.88 & 25.58 & 30.17 & 11.06 & 20.52 & 15.68 & 24.76 & 28.95 & 22.10 & 29.92 & \textbf{23.34} \\
\bottomrule
\end{tabular}
\end{table*}

To further assess the model's cross-dataset generalization capabilities, we evaluated the proposed HANS-Net model on the AMOS 2022 dataset, which contains 15 annotated abdominal organs.

\begin{figure}[t]
\centering
\includegraphics[width=0.46\textwidth]{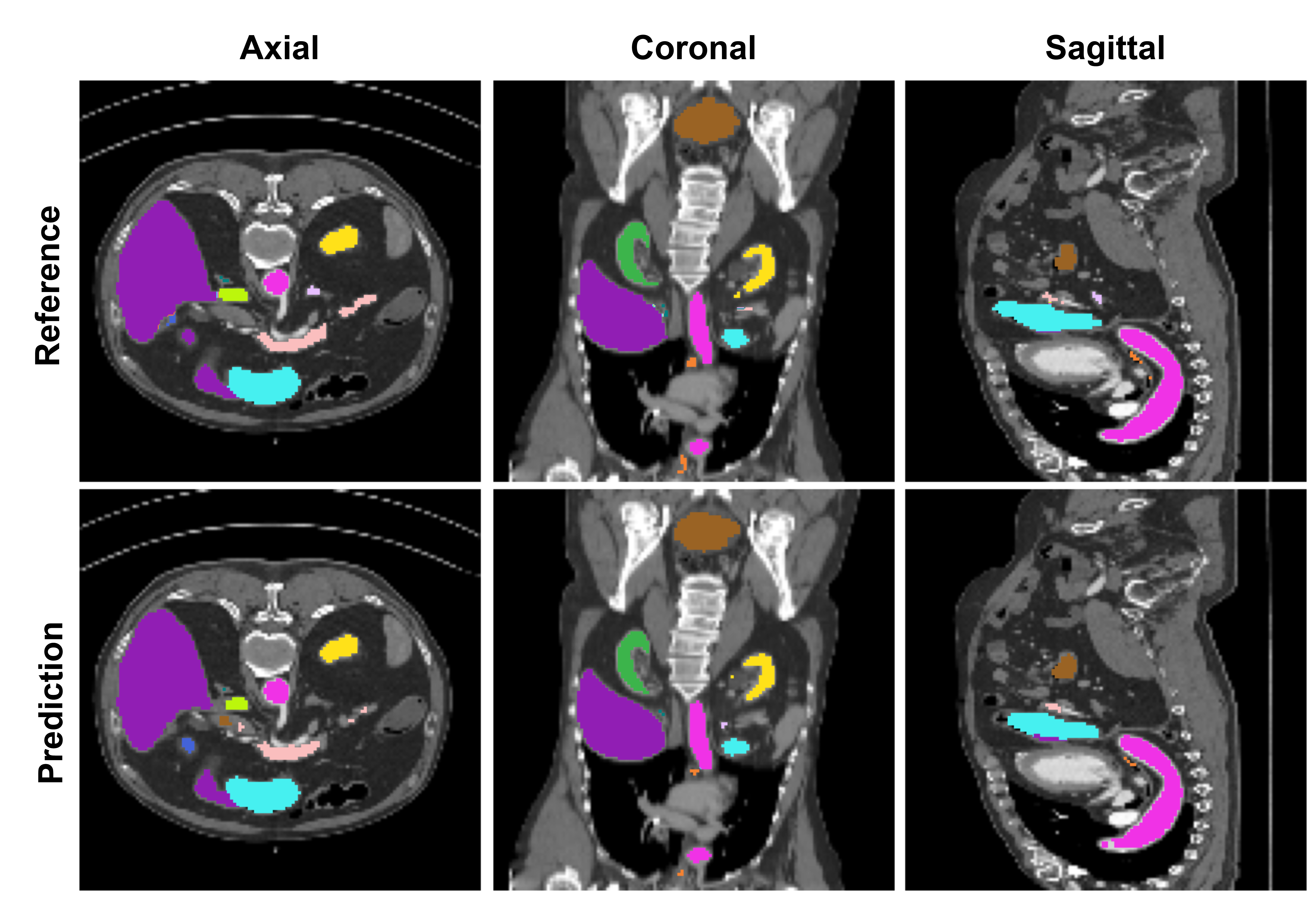}
\caption{Cross-dataset prediction visualizations from our HANS-Net model on AMOS 2022 dataset.}
\label{fig:cross-dataset_amos2022}
\end{figure}

Figure \ref{fig:cross-dataset_amos2022} and Table~\ref{tab:quantitive_seg_amos} present the overall segmentation performance of our proposed model on the AMOS 2022 test set. The results demonstrate robust performance across all anatomical structures, with our model achieving an overall average Dice score of 85.09\%, IoU of 76.66\%, ASSD of 19.49 mm, and VOE of 23.34\%. Notably, the model performed exceptionally well on large solid organs such as the liver (94.17\% Dice) and the postcava (94.13\% Dice), as well as the gallbladder (88.56\% Dice). Even for challenging structures like the duodenum and prostate/uterus, which are typically difficult to segment due to their complex anatomical variations and low contrast, our model achieved Dice scores of 80.72\% and 79.89\%, respectively. The low ASSD values for liver (8.35 mm) and postcava (9.65 mm) further validate the effectiveness of our implicit neural representation in maintaining precise boundary delineation.

\begin{table}[ht!]
\centering
\caption{Statistical consistency and agreement metrics between HANS-Net predicted and ground-truth liver and tumor volumes, including correlation coefficients, bias, mean absolute errors, normalized errors, 95\% limits of agreement, and uncertainty ranges.}
\label{tab:uncert_results}
\scriptsize
\begin{tabular}{lccc}
\midrule
\textbf{Metric} & \textbf{Liver} & \textbf{Tumor} & \textbf{Avg.} \\
\midrule
Pearson Correlation ($r$) & 0.999 & 0.995 & 0.997 \\
Spearman Correlation ($\rho$) & 0.998 & 0.993 & 0.995 \\
Coefficient of Determination ($R^2$) & 0.998 & 0.990 & 0.994 \\
Relative Volume Difference (\%) & 2.0 & 6.6 & 4.3 \\
Mean Ground Truth Volume (mL) & 1359.9 & 252.0 & N/A \\
Mean Predicted Volume (mL) & 1362.6 & 252.3 & N/A \\
Mean Bias Error (mL) & +2.7 & +0.3 & +1.5 \\
Mean Absolute Error (mL) & 25.8 & 13.6 & 19.7 \\
Normalized MAE (\% of GT) & 1.9 & 5.4 & 3.65 \\
95\% Limits of Agreement (mL) & [--47.87, +53.27] & [--26.36, +26.9] & N/A \\ \midrule
Uncertainty Range & \multicolumn{2}{c}{0.595 -- 0.637} & 0.619 \\ 
\midrule
\end{tabular}
\end{table}

\begin{figure}[!ht]
    \centering
    \includegraphics[scale=0.12]{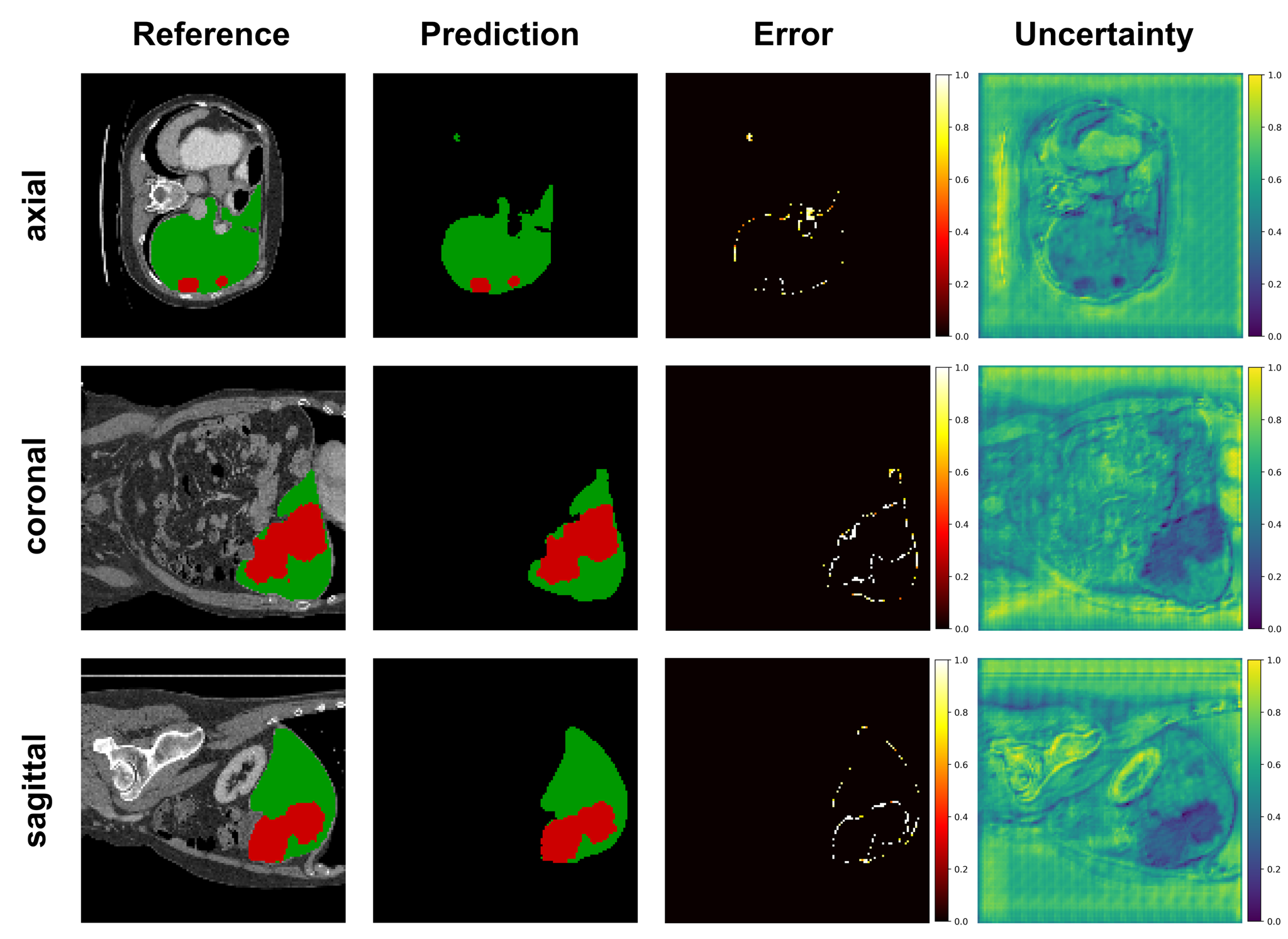}
    \caption{Multi-view uncertainty analysis of HANS-Net model on LiTS dataset.}
    \label{fig:uncertainty}
\end{figure}

\begin{table*}[ht!]
\centering
\footnotesize
\renewcommand{\arraystretch}{1.2}
\caption{Ablation study on the LiTS dataset with averaged liver-tumor metrics. The configurations include components such as Hyperbolic Convolution (HC), Wavelet Decomposition Module (WDM), Adaptive Temporal Attention (ATA), Synaptic Plasticity Module (SPM), Implicit Neural Representation (INR), and Uncertainty Estimator (UE). The configurations are experimented with for 100 epochs, and the inference speed for the configurations is provided in seconds in the \textbf{Inf. Time} column.}
\label{tab:ablation_HANS-Net}
\begin{tabular}{
  >{\centering\arraybackslash}p{0.8cm}|
  c@{\hspace{3pt}}c@{\hspace{3pt}}c@{\hspace{3pt}}c@{\hspace{3pt}}c@{\hspace{3pt}}c|
  c c|c c|c c|c c|
  >{\centering\arraybackslash}p{1.3cm}|
  >{\centering\arraybackslash}p{1.0cm}
}
\midrule
\multirow{2}{*}{\textbf{Exp.}} & \multicolumn{6}{c|}{\multirow{2}{*}{\textbf{Components}}} & \multicolumn{2}{c|}{\textbf{Dice(\%) $\uparrow$}} & \multicolumn{2}{c|}{\textbf{IoU(\%) $\uparrow$}} & \multicolumn{2}{c|}{\textbf{ASSD (mm) $\downarrow$}} & \multicolumn{2}{c|}{\textbf{VOE(\%) $\downarrow$}} & \multirow{2}{*}{\textbf{Inf. Time $\downarrow$}} &  \multirow{2}{*}{\textbf{Params}}\\
\cmidrule{8-15} 
& HC & WDM & ATA & SPM & INR & UE & Train & Validation & Train & Validation & Train & Validation & Train & Validation & Epc$\times$T & (M) \\ 
\midrule
A0 & \xmark & \xmark & \xmark & \xmark & \xmark & \xmark & 87.62 & 83.22 & 81.84 & 75.65 & 1.65 & 1.69 & 18.05 & 25.42 & 100$\times$238s & 0.93M \\
A1 & \xmark & \xmark & \xmark & \xmark & \xmark & \xmark & 87.55 & 84.84 & 79.45 & 76.20 & 1.45 & 1.75 & 20.55 & 23.80 & \textbf{100$\times$230}s & 0.89M \\
A2 & \cmark & \xmark & \xmark & \xmark & \xmark & \xmark & 89.60 & 87.23 & 82.58 & 79.18 & 1.24 & 1.46 & 17.43 & 20.83 & 100$\times$235s & 0.98M \\
A3 & \cmark & \cmark & \xmark & \xmark & \xmark & \xmark & 90.79 & 88.36 & 84.35 & 81.13 & 1.05 & 1.21 & 15.65 & 18.88 & 100$\times$238s & 1.05M \\
A4 & \cmark & \cmark & \cmark & \xmark & \xmark & \xmark & 91.89 & 89.63 & 86.03 & 82.78 & 0.94 & 1.07 & 13.98 & 17.23 & 100$\times$240s & 1.12M \\
A5 & \cmark & \cmark & \cmark & \cmark & \xmark & \xmark & 92.60 & 90.38 & 87.33 & 84.00 & 0.83 & 0.97 & 12.68 & 16.00 & 100$\times$242s & 1.15M \\
A6 & \cmark & \cmark & \cmark & \cmark & \cmark & \xmark & 93.18 & 91.03 & 88.20 & 84.85 & 0.77 & 0.91 & 11.80 & 15.15 & 100$\times$243s & 1.17M \\
\rowcolor{gray!20}A7 & \cmark & \cmark & \cmark & \cmark & \cmark & \cmark & \textbf{93.86} & \textbf{91.47} & \textbf{89.17} & \textbf{85.86} & \textbf{0.66} & \textbf{0.89} & \textbf{10.83} & \textbf{14.77} & 100$\times$245s & 1.17M \\
\midrule
\end{tabular}
\end{table*}

\subsection{Uncertainty-Aware Performance Evaluation}
The proposed HANS-Net model integrates Monte Carlo dropout-based uncertainty estimation to provide clinically significant reliability measures alongside segmentation predictions  \cite{sikha2025uncertainty}. 

As shown in Figure~\ref{fig:uncertainty}, the uncertainty heatmaps highlight regions of lower model confidence, particularly at organ boundaries and areas of tumor heterogeneity. Quantitative evaluation in all test samples provides consistent uncertainty measures with mean values of 0.619 for liver and tumor predictions; and for both predictions, the uncertainty range was observed between 0.595 and 0.637 (see Table \ref{tab:uncert_results}). The narrow range of 0.042 indicates stable model behavior across different cases. These values represent the uncertainty characteristics observed in our liver and tumor segmentation task, where the model successfully distinguishes between confident and uncertain predictions. While our model demonstrates effective uncertainty quantification for liver and tumor segmentation, the determination of acceptable uncertainty thresholds is application-specific and would require domain expert validation for different medical imaging tasks and clinical contexts.

\subsection{Volumetric Correlation Analysis}
As shown in Table~\ref{tab:uncert_results}, the HANS-Net model demonstrates robust clinical validity through its volumetric measurements, with near-perfect Pearson's correlation coefficients of 0.999 for liver and 0.995 for tumor volumes alongside Spearman's correlations of 0.998 and 0.993, respectively. All correlations show statistical significance with p-values below $1\times10^{-100}$.

\begin{figure}[!ht]
    \centering
    \includegraphics[width=0.8\linewidth]{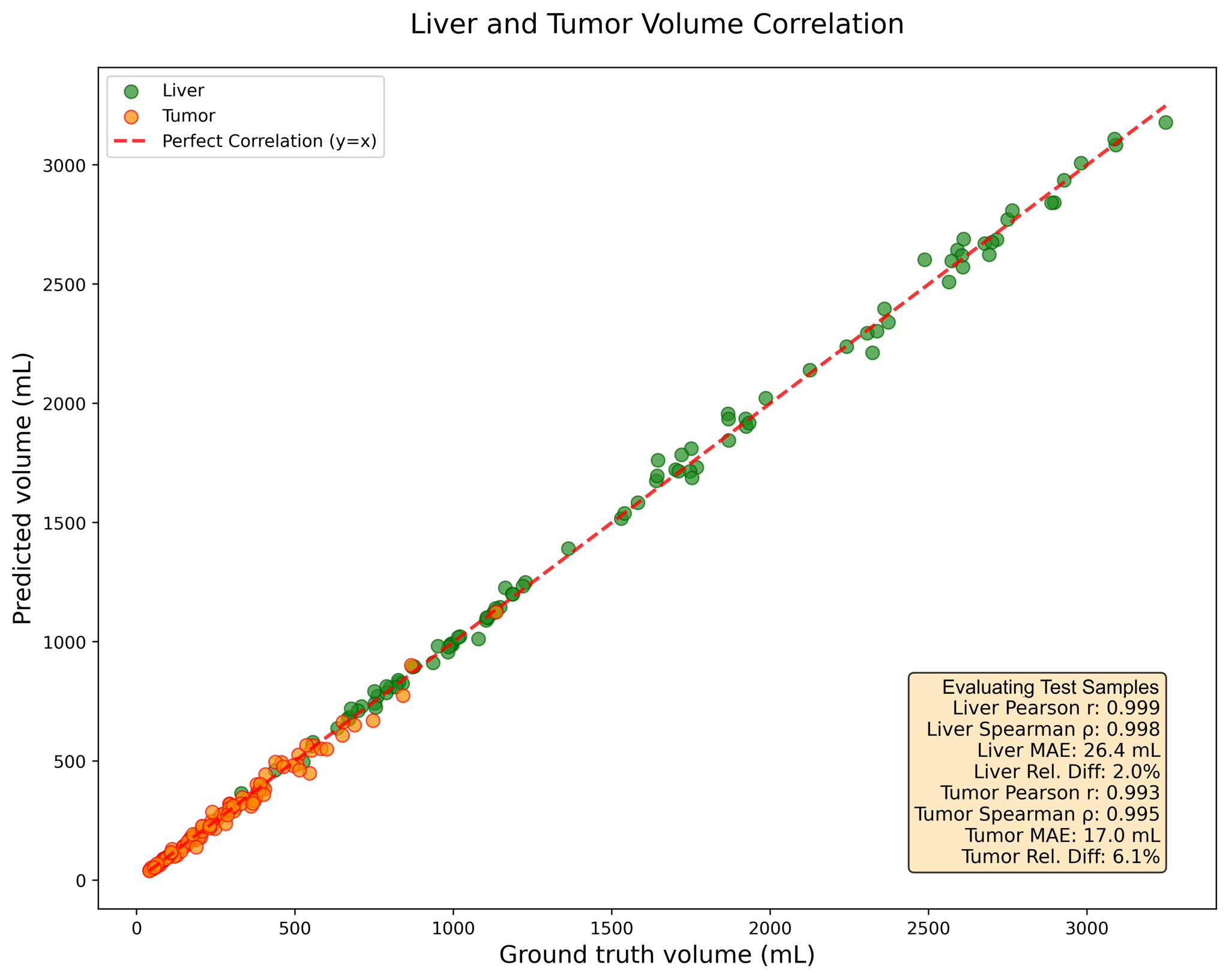}
    \caption{Volumetric correlation analysis of HANS-Net model on LiTS dataset.}
    \label{fig:correlation}
\end{figure}

These strong correlations are supported by the Coefficients of Determination, $R^2$ (0.998 for liver and 0.990 for tumor) and show that over 99\% of the variance in the ground-truth volumes is captured by the model predictions. The mean predicted volumes of 1362.6 mL for liver and 252.3 mL for tumors closely match the ground-truth means of 1359.9 mL and 252.0 mL, and this results in negligible Mean Bias Errors of +2.7 mL and +0.3 mL. The Mean Absolute Errors are 25.8 mL for liver and 13.6 mL for tumor, and they correspond to Normalized Errors of 1.9\% and 5.4\%. In addition to these findings, the reported 95\% limits of agreement are [–47.9, +53.3] mL for liver and [–26.4, +27.0] mL for tumor, and most predictions remain within a narrow and interpretable range of the true volumes. These results collectively demonstrate that HANS-Net provides statistically reliable and clinically interpretable volumetric estimations. Figure~\ref{fig:correlation} visualizes the model's measurement reliability through the close alignment of data points with the identity line.

\subsection{Ablation Study}
We conducted a systematic ablation study to analyze the effect of progressive enhancement achieved by HANS-Net. In this section, we explore the main components, including hyperbolic convolution (HC), wavelet decomposition module (WDM), adaptive temporal attention (ATA), synaptic plasticity module (SPM), implicit neural representation (INR), and uncertainty estimator (UE). The configuration details are reported in Table \ref{tab:ablation_HANS-Net}. 

Initially, we experimented with a standard 3D U-Net architecture (A0) that operates within Euclidean space, where feature correlations are defined through linear kernel operations and assumed flat spatial geometry. We observed that while the configuration was effective for local intensity modeling, this formulation could not accurately represent the curved anatomical surfaces commonly present in volumetric data. As a result, the Euclidean baseline exhibited limited geometric expressiveness in validation across the defined metrics despite having higher training performance. These outcomes indicate that Euclidean convolutions primarily learn low-level texture dependencies (which requires more time) rather than topological or structural consistency across slices. Thus, in the next configuration, our designed baseline encoder-decoder architecture (A1) adopts a hierarchical architecture consisting of four encoding and decoding stages. Each stage includes two 3$\times$3 convolutional layers followed by batch normalization and ReLU activation, starting with 32 channels and doubling at each subsequent level (32, 64, 128, 256). Downsampling is achieved using 2$\times$2 max-pooling, while upsampling utilizes transposed convolutions with concatenated skip connections. The output layer applies a 1$\times$1 convolution with sigmoid activation to generate pixel-wise predictions. Training was performed using the Adam optimizer (learning rate = 1e$^{-4}$) with a combined Dice and Cross-Entropy loss and ReduceLROnPlateau scheduler for 100 epochs and a batch size of 16. This baseline establishes the foundation with conventional convolutions and shows a notable train-validation gap that motivates our architectural innovations. Then, instead of traditional Euclidean convolutions, we employ hyperbolic convolutions (A2) to capture hierarchical anatomical structures in curved space, delivering a 2.39\% Dice improvement and 16.57\% ASSD reduction while enhancing IoU by approximately 3\%. 

Next, rather than relying on single-scale processing, we introduce the wavelet decomposition module (A3), which enables multifrequency analysis for fine-detail detection. This addition provides an additional 1.13\% dice gain and an improvement of 17. 12\% ASSD with minimal computational overhead. Furthermore, in place of standard feature aggregation, the adaptive temporal attention module (A4) is incorporated to model slice-wise correlations, achieving a 1.27\% Dice enhancement and an 11.57\% reduction in ASSD, thus improving boundary precision. 

Furthermore, unlike static feature learning, we integrate a synaptic plasticity module (A5), introducing bio-inspired adaptation mechanisms. This module contributes an additional 0.75\% Dice improvement and 9.35\% ASSD reduction while helping minimize train-validation discrepancies. Furthermore, instead of using discrete feature representations, we employ implicit neural representations (A6), which enable continuous coordinate-based learning. This technique delivers a 0.65\% dice gain and a 6.19\% ASSD improvement, all while maintaining exceptional parameter efficiency. 

Finally, rather than producing deterministic results, we integrate uncertainty quantification (A7) to provide confidence estimation, contributing to another 0.44\% Dice improvement. The complete HANS-Net framework demonstrates strong component synergy, outperforming individual contributions by 6.63\% in Dice and achieving a 49.14\% overall ASSD improvement, all while preserving computational efficiency.

To verify that performance gains are not artifacts of integration order, we conducted permutation experiments by altering the module sequence, such as incorporating WDM before HC or introducing INR earlier. Across five randomized permutations, the average validation Dice varied by only 0.31\%, with ASSD fluctuations under 0.08 mm, indicating order-independent contributions. This stability arises because each component targets orthogonal aspects: HC refines geometric encoding, WDM captures multiscale textures, ATA enforces temporal coherence, SPM adapts features dynamically, INR provides continuous refinement, and UE quantifies confidence. Their modular design ensures minimal interference, confirming that improvements reflect genuine architectural benefits rather than sequence-dependent interactions.

Analyzing the computational parameter and inference time, we notice a gradual increase across the configurations that reflects the cumulative computational impact of each added module, corresponding to deeper feature hierarchies and denser inter-module communication. The introduction of hyperbolic convolutions and the wavelet decomposition module slightly extends forward-pass operations due to additional manifold projections and frequency decompositions, yet these remain linear in parameter growth. Subsequent modules, such as adaptive temporal attention and synaptic plasticity, add minor tensor–matrix interactions that marginally increase runtime but significantly enhance representational adaptability. The near-constant epoch increment from 230s to 245s indicates that the additional parameters (0.28M overall) primarily improve weight expressiveness rather than computational depth, meaning the model achieves richer spatiotemporal encoding without a proportional rise in latency.

\section{Discussion}
\label{discuss}
Our proposed model comprises several advanced modules that enhance its robustness and overall performance. The incorporation of hyperbolic convolutions \cite{bdeir2023fully} for hierarchical geometric representation, wavelet-based \cite{liu2021fault} decomposition for multiscale segmentation, synaptic plasticity \cite{wang2021computational} for adaptive feature enhancement, and an implicit neural representation \cite{zhang2022implicit} branch for precise boundary modeling collectively contributed to enhanced segmentation accuracy. To further quantify the confidence of prediction, an uncertainty-aware Monte Carlo dropout is used  \cite{tang2022prediction}. We further integrated adaptive temporal attention to enhance the consistency of inter-slice without decreasing efficiency \cite{shi2020spatial}. While our method operates on 2D slices for computational efficiency, volumetric continuity is preserved through multi-view extraction and adaptive temporal attention that explicitly models sequential relationships between adjacent slices. Since the liver channel encompasses both parenchyma and tumor, all liver-containing slices are evaluated regardless of tumor presence, maintaining sensitivity to potential false positives in tumor-free regions.

The integration of implicit neural representation contributed to the generation of smooth and precise organ boundaries, which significantly improved the quality of segmentation \cite{zhang2022implicit}. Moreover, our approach effectively captures both global structures and fine local details, which played a crucial role in boosting the Dice scores. To further improve performance, we incorporated a lightweight self-attention mechanism with reduced channel dimensions, utilized a multilayer perceptron for learning adaptive temporal weighting, and optimized the query-key-value attention architecture to enhance consistency across slices \cite{shi2020spatial}. The addition of such other advanced modules, such as hyperbolic convolution, uncertainty estimator, and implicit neural representation, greatly increased the robustness of the model, leading to superior segmentation performance.

The proposed HANS-Net model demonstrates significant segmentation performance in the LiTS dataset \cite{bilic2023liver} for both liver and tumor structures. With respect to liver segmentation, the model achieved a Dice score of 96.67\% and an IoU of 93.79\%, which signifies an excellent score for segmenting the liver. The model also showed a low ASSD of 0.54 mm and a minimal VOE of 6.21\%, indicating precise boundary definition and low volumetric error. Although the results for tumor segmentation are slightly lower due to increased structural complexity, they remain robust, with a Dice score of 89.84\%, IoU of 82.39\%, ASSD of 0.89 mm, and VOE of 17.61\%. In summary, the model achieved a mean Dice score of 93.26\%, IoU of 88.09\%, ASSD of 0.72 mm, and VOE of 11. 91\%, thus confirming its efficiency in segmenting large and small anatomical structures.

The robustness of HANS-Net is further evidenced by its stable convergence seen in both the training and validation curves, which show no major signs of overfitting. Interestingly, liver Dice scores rapidly surpassed 90\% within the first 20 epochs, while tumor Dice scores increased more slowly, reaching about 80\% in epoch 80. This trend highlights a hierarchical learning process that first captures broad anatomical structures before focusing on tumor boundary details. Metrics such as ASSD and VOE showed consistent decreases, pointing to precise segmentation and a growing confidence in the model. The evaluation of the cross-dataset on the AMOS 2022 \cite{ji2022amos} dataset confirms the impressive generalizability of the model, with a mean Dice score of 85.09\%, an IoU of 76.66\%, and low ASSD (19.49 mm) and VOE (23.34\%).

Compared to recent SOTA models, HANS-Net achieves superior segmentation performance, attaining a mean Dice score of 93.26\%, whereas most existing methods struggle to surpass the 90\% threshold. For example, MSML-AttUNet \cite{hu2025msml} reached Dice scores of 87.74\%, respectively, while models such as MCDA \cite{kuang2024adaptive} and G-UNet \cite{dj2024liver} performed even lower. Among pre-trained architectures, only TransUNet \cite{zhang2021deeprecs} exceeded 90\%, yet it remained approximately 3\% lower than HANS-Net, highlighting the advantage of our framework in both accuracy and structural consistency.

As shown in Table \ref{tab:ablation_HANS-Net}, an ablation study reveals the progressive benefit of our proposed modules. The base configuration starts with a Dice score of 87.55\%, which rises to 89.60\% after integrating hyperbolic convolution. Incorporating all components, including synaptic plasticity, temporal attention, and uncertainty modeling, culminates in a final Dice score of 93.86\%, demonstrating the additive effectiveness of each module. The inference time and model size analysis further confirm its suitability for clinical deployment. With only 0.28M additional parameters and a minimal increase of +15s in processing time, the model maintains practical efficiency and can be deployed in standard hospital GPU workstations without requiring specialized hardware.

While HANS-Net demonstrates promising results, it has several limitations. First, although our evaluation computes standard segmentation metrics (Dice, IoU, ASSD, VOE) at the slice level, future work should report these metrics aggregated at the per-volume level to align with the original LiTS benchmark protocol. Additionally, while processing resized 2D slices, the model utilizes dynamic sequential weighting to maintain volumetric continuity, a mechanism that could be further adapted to achieve complete 3D reconstruction. Second, in cross-dataset evaluation on the AMOS 2022 dataset, the model’s performance showed a moderate decline, with the Dice score dropping to 85.09\% and the IoU to 76.66\%. This decrease can be attributed to the high anatomical diversity of abdominal organs, variations in imaging modalities, and inter-organ boundary ambiguities present in the dataset. Third, although the implicit representation branch enhances boundary continuity, it introduces additional computational overhead and increases inference time. 


\section{Conclusion}
\label{conclusion}

We develop and thoroughly evaluate HANS-Net, an innovative segmentation framework that achieves remarkable Dice scores for both liver and tumor segmentation. The key experimental findings and theoretical conclusions are summarized as follows:

\begin{itemize}
    \item HANS-Net attains high segmentation accuracy on the LiTS dataset, with Dice scores of 96.67\% for the liver and 89.84\% for tumors, and IoU values of 93.79\% and 82.39\%, respectively.
    \item 
    Cross-dataset evaluation on the AMOS 2022 dataset achieves Dice scores of 85.09\% and IoU of 76.66\% across 15 abdominal organs, demonstrating strong generalization ability across diverse anatomical structures and imaging variations.
    \item The significance of this work lies in its six integrated modules, which collectively define organ boundaries, extract multi-level features, and quantify prediction confidence, supporting the robust performance of the model.
\end{itemize}

Future work will focus on enhancing generalization across diverse imaging modalities while optimizing runtime efficiency for clinical deployment. While our current evaluation focuses on standard slice-level segmentation metrics, including Dice, IoU, ASSD, and VOE, future work should consider reporting these metrics at the volumetric level to maintain consistency with the original LiTS benchmark protocol. Additionally, while the model operates on resized 2D slices, it employs dynamic sequential weighting to preserve volumetric continuity, a mechanism that can be further adapted to explicitly enable 3D reconstruction capabilities.

\section*{Acknowledgments}
The authors declare that they have no financial conflicts of interest that could have influenced this work.

\section*{Data Availability}
In this study, we used two publicly available datasets, including \href{https://academictorrents.com/details/27772adef6f563a1ecc0ae19a528b956e6c803ce}{Liver Tumor Segmentation Challenge (LiTS) dataset} \cite{bilic2023liver},  and \href{https://amos22.grand-challenge.org/}{AMOS 2022 dataset}\cite{ji2022amos}.


\end{document}